\documentclass[11pt]{article}
\usepackage{amsmath,amssymb,amsthm,subfig}
\usepackage[mathscr]{eucal}
\usepackage{rotate,graphics,epsfig}
\usepackage{multirow}

 \newcommand{\field}[1]{\mathbb{#1}}
 \newcommand{\R}{\field{R}}
\newcommand{\C}{\field{C}}
 \newcommand{\Exp}{\field{E}}
 \newcommand{\Pro}{\field{P}}
\newcommand{\Z}{\field{Z}}
\newcommand{\df}{\,\mathrm{d}}

\newcommand{\be}{\begin{eqnarray}}
\newcommand{\ee}{\end{eqnarray}}
\newcommand{\by}{\begin{eqnarray*}}
\newcommand{\ey}{\end{eqnarray*}}
\newcommand{\bn}{\begin{enumerate}}
\newcommand{\en}{\end{enumerate}}
\newcommand{\bi}{\begin{itemize}}
\newcommand{\ei}{\end{itemize}}

\newtheorem{theo}{Theorem}[section]
\newtheorem{pr}{Proposition}[section]
\newtheorem{lem}{Lemma}[section]
\newtheorem{co}{Corollary}[section]
\newtheorem{re}{Remark}[section]
\newtheorem{de}{Definition}[section]
\newtheorem{exa}{Example}[section]

\newcommand{\bex}{\begin{exa}}
\newcommand{\eex}{\end{exa}}
\newcommand{\bt}{\begin{theo}}
\newcommand{\et}{\end{theo}}
\newcommand{\bp}{\begin{pr}}
\newcommand{\ep}{\end{pr}}
\newcommand{\bl}{\begin{lem}}
\newcommand{\el}{\end{lem}}
\newcommand{\bc}{\begin{co}}
\newcommand{\ec}{\end{co}}
\newcommand{\br}{\begin{re}}
\newcommand{\er}{\end{re}}
\newcommand{\bd}{\begin{de}}
\newcommand{\ed}{\end{de}}

\numberwithin{equation}{section}
\newcommand{\nb}{\nonumber}
\newcommand{\ol}{\overline}
\newcommand{\td}{\tilde}

\newbox\tmp
\newdimen\height
\newdimen\dropdist
\def\BX#1{\setbox\tmp=\hbox{$\overline{\scriptstyle #1}$}
              \height=\ht\tmp
              \dropdist=\dp\tmp
              \advance\dropdist by .7pt
              \advance\height by \dp\tmp
              \box\tmp
              \lower \dropdist \hbox{\vrule height
              \height width .25pt\relax}
              \ifnum0=`{\else}\fi
}
\def\bx{\expandafter\BX\expandafter{\ifnum0=`}\fi}
\usepackage[sort]{natbib}
\begin{document}
\title{An Identity of Hitting Times and Its Application to the Valuation of Guaranteed Minimum Withdrawal Benefit}
\author{
    Runhuan Feng\\
    Department of Mathematics\\
    University of Illinois at Urbana-Champaign\\
    rfeng@illinois.edu
  \and
    Hans W. Volkmer\\
    Department of Mathematical Sciences\\
    University of Wisconsin - Milwaukee\\
    volkmer@uwm.edu
}
\maketitle

\begin{abstract}

In this paper we explore an identity in distribution of hitting times of a finite variation process (Yor's process) and a diffusion process (geometric Brownian motion with affine drift), which arise from various applications in financial mathematics. As a result, we provide analytical solutions to the fair charge of variable annuity guaranteed minimum withdrawal benefit(GMWB) from a policyholder's point of view, which was only previously obtained in the literature by numerical methods. We also use complex inversion methods to derive analytical solutions to the fair charge of the GMWB from an insurer's point of view, which is used in the market practice, however, based on Monte Carlo simulations. Despite of their seemingly different formulations, we can prove under certain assumptions the two pricing approaches are equivalent.


\end{abstract}

\section{Introduction}

There are two sets of stochastic processes that are of particular interests for financial and actuarial applications. One of them is the time-integral of a geometric Brownian motion, defined by
\[A^{(\nu)}_t:=\int^t_0 \exp\{2 B^{(\nu)}_u\} \df u,\qquad B^{(\nu)}_t:=B_t+ \nu t ,\] where $B$ is a standard Brownian motion.  It is also known as Yor's process in computational finance literature and arises from the pricing of continuously monitoring Asian (average price) options. Detailed accounts of the laws of the geometric Brownian motion and its time-integral as well as their applications in mathematical finance can be found in \cite{Yor92}, \cite{GemYor}, \cite{Yor01}, \cite{CarSch}, etc. 

The other one is a set of diffusion processes, known as geometric Brownian motions with affine drift, which are defined as solutions to stochastic differential equations for $t>0,$
\be \df X_t&=&[2(\nu+1) X_t+1 ] \df t+ 2 X_t \df B_t,\quad X_0=x. \label{Isde}\\
\df Y_t&=& [2(\nu+1) Y_t-1] \df t +2 Y_t \df B_t,\qquad Y_0=y. \label{Dsde}
\ee
It follows immediately from It\^o's formula that for $t\ge 0,$
\be X_t&=& \exp\{2B^{(\nu)}_t\} \left( x+\int^t_0 \exp\{-2B^{(\nu)}_s\} \df s\right),\nb\\
 Y_t&=&\exp\{2B^{(\nu)}_t\} \left( y-\int^t_0 \exp\{-2B^{(\nu)}_s\} \df s\right).\label{D}\ee From time to time, we also use the notation $X^{(\nu)}$ and $Y^{(\nu)}$ to indicate the parameter used in their definitions in comparison with the process $A^{(\nu)}$. The process $X$ was introduced for financial applications in \cite{Lew} for the pricing of European-style options on dividend paying stocks. It is also well-known using a duality lemma of L\'evy process (c.f. \cite[Lemma 3.4]{Kyp}) that for $x=0$ and each fixed $t\ge 0$,
\be  X^{(\nu)}_t \sim  A^{(\nu)}_{t}\, , \label{IA}\ee where $\sim$ means ``equals in distribution" throughout the note. This identity in distribution is exploited extensively in many papers for alternative methods for the pricing of Asian options, such as \cite{DonGhoYor}, \cite{Lin04b}, etc. There appears to be little discussion about the process $Y$ in the existing literature.

\begin{figure}[htb]
  \centering
  \subfloat[Sample paths of $A^{(\nu)}$]{\includegraphics[width=2.5in,height=2.5in]{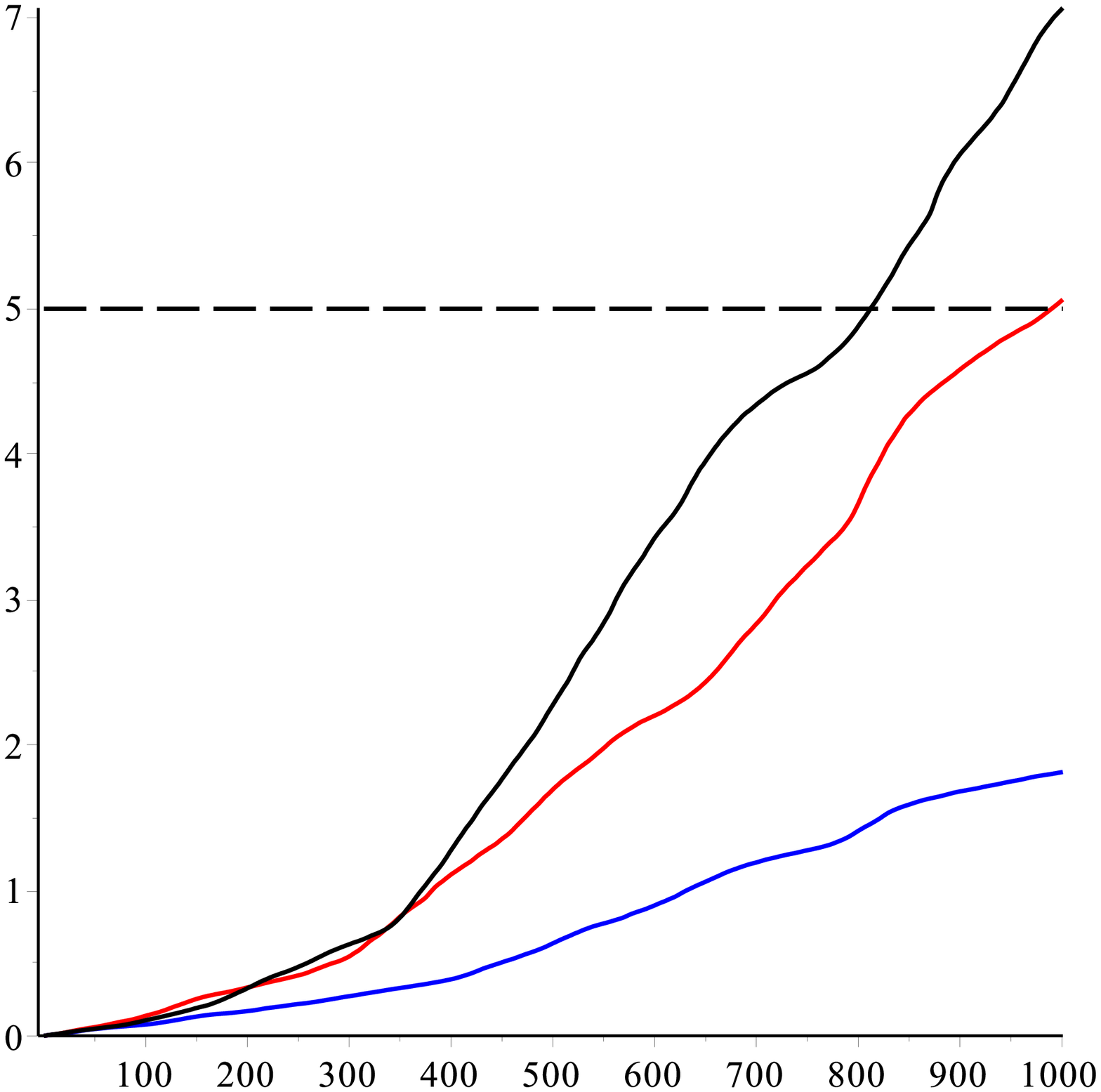}}
  \hspace{0.5in}
  \subfloat[Sample paths of $Y^{(-\nu)}$]{\includegraphics[width=2.5in,height=2.5in]{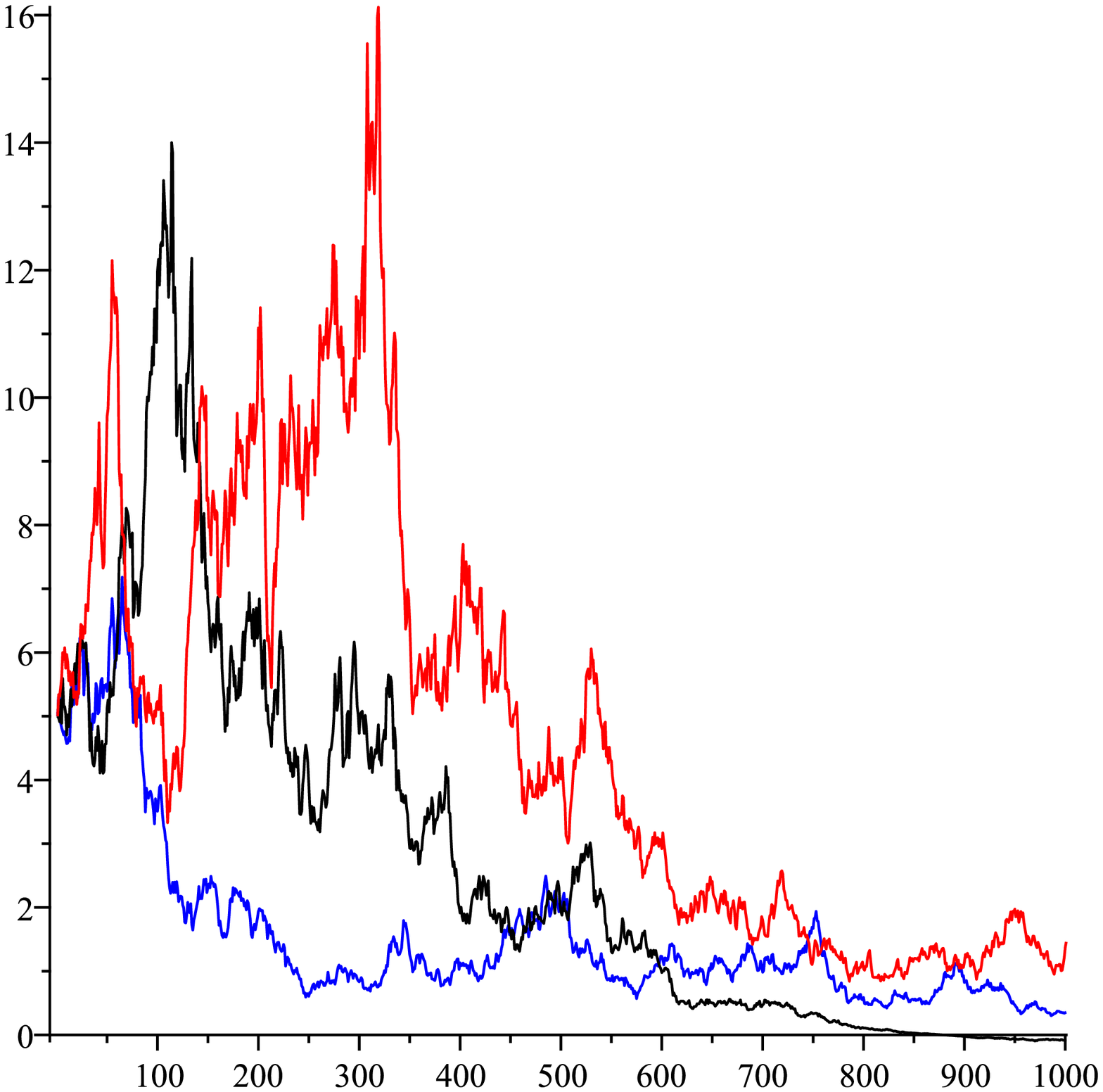}}
  % if separate labels are needed, use {\label{figurename} \includegraphics[]{}}
  \caption{Comparison of sample paths: $t=1, \nu=1, x=5.$ Horizontal axis $1000t$.}\label{fig1}
\end{figure}

In this paper, we investigate the first passage times of the processes $A^{(\nu)}$ and $Y^{(\nu)}$. Let
\[H^{(\nu)}_x=\inf\{t: A^{(\nu)}_t =x\},\qquad \tau^{(\nu)}_{x,y}=\inf\{t: Y^{(\nu)}_0=x, Y^{(\nu)}_t=y\},\qquad x \ge 0.\] We shall also write $H_x=H^{(\nu)}_x$ and $\tau_{x,y}=\tau^{(\nu)}_{x,y}$ for short when the parameter $\nu$ is known from the context. It is immediately clear from \eqref{D} that
\be \tau^{(-\nu)}_{x,0}\sim  H^{(\nu)}_{x}.\label{tauDA}\ee This appears to be a rather peculiar identity, which roughly means that for every sample path of the process $A^{(\nu)}$ starting off at $0$ and ascending to $x$, there is a sample path of the process $Y^{(-\nu)}$ starting off at $x$ that takes just as much time descending (not monotonically) to $0$. Although the two processes are intricately connected by the hitting times, they have drastically different sample path properties as evident from Figure \ref{fig1}. For instance, the process $A^{(\nu)}$ is a process of finite variation whereas the process $Y$ has infinite variation; the process $A^{(\nu)}$ is an almost surely increasing process while no such monotonicity can be said of the diffusion process $Y$.

%It was shown using the duality of Levy process that they have the same finite dimensional distribution when the parameters are properly chosen,
% $\mu=2(\nu+1)$ and $\sigma=2$.

In Sections 2 and 3, we analyze the two types of hitting times seperately using their distinct analytic properties. As an application of the identity in distribution \eqref{tauDA} in Section 4, we develop closed-form solutions to various quantities required for the valuation of the guaranteed minimum withdrawal benefit (GMWB).

\section{Hitting time of Yor's process}

The distribution of $H_a, a>0$ was one of many computational issues raised but not directly addressed in \cite[Section 8.2]{Yor01b} regarding the exponential functionals of Brownian motion in connection with Asian options. In this section, we derive two equivalent expressions for the distribution of $H_a$. Since we rely on a key result from \cite{Yor92} as well as the transition density function of Bessel process with positive index, it is assumed throughout this section that $\nu \ge 0$.

\bp \label{prop:LapHa} For $\nu\ge 0$, the Laplace transform of $H_a$ is given by
\be \Exp\left[ e^{ -s H_a }  \right]=(2a)^{(\nu-\lambda)/2} \exp \left(-\frac{1}{2a}  \right)\frac{\Gamma((\nu+\lambda)/2+1)}{\Gamma(\lambda+1)} M\left(\frac{1}{2}(\nu+\lambda)+1, \lambda+1; \frac{1}{2a}\right), \quad\label{Higbm}\ee
where  $\lambda=\sqrt{2s+\nu^2}$ and $M$ is the Kummer's function of the first kind.
\ep

\begin{proof}
It is shown in \cite{Yor92} that for $\nu\ge 0$
\[\Exp \left[ \left.\exp \left( -\frac{\theta^2}{2} H_a \right) \right| R^{(\nu)}_a =ra \right]=\frac{I_\lambda(r)}{I_{\nu}(r)},\quad \lambda=\sqrt{\theta^2+\nu^2}.\]
The known density of the Bessel process $R^{(\nu)}$ is, for $\nu\ge 0,$
\[p_t(\rho)=\frac{\rho^{\nu+1}}{t} \exp \left(-\frac{1}{2t}(1+\rho^2) \right) I_\nu \left( \frac{\rho}{t}  \right). \]
Therefore, we can find the Laplace transform of $H_a$ by
\be \Exp \left[ \exp \left( -\frac{\theta^2}{2} H_a \right) \right]
&=& \int^\infty_0 \Exp \left[ \left.\exp \left( -\frac{\theta^2}{2} H_a \right) \right| R^{(\nu)}_a =\rho \right]p_a(\rho) \df \rho\nb\\
&=&\int^\infty_0 I_\lambda\left(\frac{\rho}{a}\right)\frac{\rho^{\nu+1}}{a}\exp\left( -\frac{1}{2a}(1+\rho^2)\right) \df \rho.\label{lap}\ee
It follows from \cite[page 383]{Wat} that
\by &&\int^\infty_0 I_\lambda\left(\frac{\rho}{a}\right)\frac{\rho^{\nu+1}}{a}\exp\left( -\frac{1}{2a}(1+\rho^2)\right) \df \rho\\
&=& (2a)^{(\nu-\lambda)/2} \exp \left(-\frac{1}{2a}  \right)\frac{\Gamma((\nu+\lambda)/2+1)}{\Gamma(\lambda+1)} M\left(\frac{1}{2}(\nu+\lambda)+1, \lambda+1; \frac{1}{2a}\right).\ey

\end{proof}

\bp {\bf (First Representation)} For $\nu \ge 0$,
the probability density of $H_a$ is given by
\be f(u)=\exp\left(-\frac12\nu^2u-\frac1{2a}+\frac{\pi^2}{2u}\right)(2\pi^3u)^{-1/2}a^{\nu+1}
\int_0^\infty \exp(-\frac{y^2}{2u})\sinh y\sin\left(\frac{\pi y}{u}\right) g(y)\,dy, \label{first}
\ee where \be g(y)=a^{-\frac12\nu-\frac32}\Gamma(\nu+3)\exp\left(\frac1{4a}\cosh^2y\right)U(\nu+\tfrac52,a^{-1/2}\cosh y),\qquad \label{g}\ee and $U(b,z)$ is a parabolic cylinder function related to Kummer's function of second kind by
\[ U(\tfrac12b+\tfrac14,\tfrac12,\tfrac{z^2}{2})=2^{\frac12b+\frac14}e^{\frac14z^2}U(b,z) .\]
\ep
\begin{proof}
It is known from Yor(1992) that
 \[ I_{|\nu|}(r)=\int^\infty_0 \exp \left( -\frac{\nu^2 u}{2} \right) \theta_ r(u)\df u, \] where
 \[\theta_r(u)=\frac{r}{(2\pi^3 u)^{1/2}}\exp\left( \frac{\pi^2}{2u}\right) \psi_r(u),\qquad \psi_r(u)=\int^\infty_0 \exp \left(-\frac{y^2}{2u}  \right)\exp(-r\cosh y) \sinh y \sin \left( \frac{\pi y}{u} \right) \df y.\] Therefore,
 \by \Exp\left[ \exp \left( -\frac{\theta^2}{2} H_a \right) \right]&=&\int^\infty_0 \int^\infty_0 \exp \left( -\frac{(\theta^2+\nu^2) u}{2} \right)\theta_{\rho/a}(u)\df u\frac{\rho^{\nu+1}}{a}\exp\left( -\frac{1}{2a}(1+\rho^2)\right) \df \rho\\
 &=&\int^\infty_0 \exp\left( -\frac{\theta^2 u}{2} \right) \int^\infty_0 \exp \left( -\frac{\nu^2 u}{2} \right)\frac{\rho^{\nu+1}}{a}\exp\left( -\frac{1}{2a}(1+\rho^2)\right) \theta_{\rho/a}(u) \df \rho\, \df u.\ey
 Due to the one-to-one correspondence of Laplace transform and a continuous density function, we must have the density function $f(u)$ of the hitting time $H_a$
%\[ f(u)=\exp \left( -\frac{\nu^2 u}{2} \right) \int^\infty_0 \frac{\rho^{\nu+1}}{a}\exp\left( -\frac{1}{2a}(1+\rho^2)\right) \theta_{\rho/a}(u) \df \rho, \]
%or
\[ f(u)=\exp \left( -\frac{\nu^2 u}{2} \right) \int^\infty_0 (ax)^{\nu+1}\exp\left( -\frac{1}{2a}(1+a^2 x^2)\right) \theta_{x}(u) \df x. \]
Denote the integrand by $g(x,y)$. On one hand, note that there exists some $M_1>0$ such that
\[|\theta_x(u)|\le M_1 x \int^\infty_0 e^{-x e^y/2} \df y=M_1x E_1(x/2).\] Using the asymptotics of $E_1$ \cite[p.153,(6.12.1)]{NIST}, we know that for some $M_2>0$ as $x \rightarrow \infty$.
\[x^{\nu+2}\exp\left( -\frac{1}{2a}(1+a^2 x^2)\right)E_1(\frac x2)\sim M_2 x^{\nu+1}\exp\left(-\frac{ax^2+x}{2} \right).\] Similarly, using \cite[p.151,(6.6.2)]{NIST}, we know that for some $M_3<0$ as $x \rightarrow 0+$,
\[x^{\nu+2}\exp\left( -\frac{1}{2a}(1+a^2 x^2)\right)E_1(\frac x2)\sim M_3 x^{\nu+2}\ln x,\] which approaches zero by \cite[p.107,(4.4.14)]{NIST}. Thus $\int^\infty_0 \int^\infty_0 |g(x,y)|\df y \df x<\infty.$ %On the other hand, for some $M_4>0$
%\[\int^\infty_0 x^{\nu+2} e^{-\frac12 (a+e^y)x} \df x = M_4 ( a+e^y )^{-\nu-3}.\] Thus for some %$M_5>0$, as $y\rightarrow \infty$,
%\[\int^\infty_0 |g(x,y)| \df x \sim M_5  ( a+e^y )^{-\nu-3} \exp\left\{ -\frac{y^2}{2u} \right\}.\] Therefore, %$\int^\infty_0 \int^\infty_0 |g(x,y)|\df x \df y<\infty$.
By Fubini's theorem,
we exchange the integrals and obtain
\[ f(u)=\exp\left(-\frac12v^2u-\frac1{2a}+\frac{\pi^2}{2u}\right)(2\pi^3u)^{-1/2}a^{\nu+1}
\int_0^\infty \exp(-\frac{y^2}{2u})\sinh y\sin\left(\frac{\pi y}{u}\right) g(y)\,dy,
\] where \[
g(y)=\int_0^\infty x^{\nu+2}\exp\left(-\frac{ax^2}{2}-x\cosh y\right)\,dx.
\]
According to \cite[page 208]{Olv}, the following formula holds
\[ U(b,z)=\frac{\exp(-\frac14z^2)}{\Gamma(b+\frac12)}\int_0^\infty \exp\left(-zt-\frac12t^2\right)t^{b-\frac12}\,dt ,\quad b>-\frac12.
\]
Assuming $a>0$ and substituting $t=\sqrt{a}x$ we obtain
\[ U(b,z)=\frac{\exp(-\frac14 z^2)}{\Gamma(b+\frac12)}a^{\frac12b+\frac14} \int_0^\infty \exp\left(-z\sqrt{a} x-\frac12a x^2\right)x^{b-\frac12}\, dx .\]
We choose $b=\nu+\frac52$ and $z=a^{-1/2}\cosh y$ and obtain \eqref{g}.
\end{proof}

\bp {\bf (Second Representation)} For $\nu\ge 0$, the probability density of $H_a$ is given by
\be f(u)=\frac1{2\pi^2}(2a)^{(\nu+1)/2}e^{-\frac1{4a}}\int^\infty_{0}e^{-(\nu^2+p^2)u/2}W_{-(\nu+1)/2,ip/2}(\frac1{2a})\left|\Gamma\left(1+\frac{\nu+ip}{2}\right)\right|^2 \sinh(\pi p) p \df p, \;\;\;
\label{second}\ee where $W$ is the Whittaker function of the second kind.
\ep 
\begin{proof}
The distribution function of $X^{(\nu)}_t$ with $\nu\ge 0$ is obtained in \cite[(3.26)]{FenVol2}. Using the indentity in distribution \eqref{IA}, we can obtain the distribution of $A^{(\nu)}_t$ by letting $wx_0=a$ and $x_0 \rightarrow 0+$ in \cite[(3.26)]{FenVol2},
\be
\Pro(A^{(\nu)}_t <a)=\frac1{4\pi^2}(2a)^{(\nu+1)/2}e^{-\frac1{4a}}\int^\infty_{0}e^{-(\nu^2+p^2)t/2}W_{-(\nu+1)/2,ip/2}(\frac1{2a})\left|\Gamma(\frac{\nu+ip}{2})\right|^2 \sinh(\pi p) p \df p. \quad \label{A}
\ee We note that
\be \Pro(H_a>t)=\Pro(A^{(\nu)}_t<a).\label{HA}\ee Therefore, differentiating \eqref{A} w.r.t. $t$ and taking the opposite sign yields the density \eqref{second}.
\end{proof}

\br
We can demonstrate directly that the two representations  of the hitting time density are indeed equivalent. Let $\td{f}(s)$ be the Laplace transform of $f$ in \eqref{second} and $\td{P}$ be the Laplace transform of the distribution function of $A^{(\nu)}_t$ w.r.t. $t$. We can find $\td{P}$ by setting $wx_0=a$ and letting $x_0 \rightarrow 0+$ in \cite[(4.6)]{FenVol2} that
\[\td{P}(s)=\frac1s- \frac{\Gamma(\mu-\kappa+1/2)}{\Gamma(1+2\mu)}\frac{a^{1-\kappa}2^{-\kappa}}{\mu+\kappa-1/2}\exp(-\frac1{4a})M_{\kappa-1,\mu}(\frac1{2a}),\] where $\kappa=(1-\nu)/2, \mu=\sqrt{\nu^2+2s}/2.$ Using \cite[(13.14.2)]{NIST},
\[M_{\kappa,\mu}(z)=e^{-\frac12 z} z^{\frac12 +\mu} M(\frac12+\mu-\kappa,1+2\mu,z),\]
we obtain
\[\td{P}(s)=\frac1s-\frac{\Gamma(\mu-\kappa+1/2)}{\Gamma(1+2\mu)}\frac{a^{\frac12-\mu-\kappa}2^{-\frac12-\mu-\kappa}}{\mu+\kappa-1/2}\exp(-\frac1{2a})M\left(\frac32+\mu-\kappa,1+2\mu,\frac1{2a}\right).\]
 It follows from \eqref{HA} that $\td{f}(s)=1-s\td{P}(s).$ Thus,
\[\td{f}(s)=s\frac{\Gamma(\mu-\kappa+1/2)}{\Gamma(1+2\mu)}\frac{a^{\frac12-\mu-\kappa}2^{-\frac12-\mu-\kappa}}{\mu+\kappa-1/2}\exp(-\frac1{2a})M\left(\frac32+\mu-\kappa,1+2\mu,\frac1{2a}\right).\]
This agrees with the Laplace transform \eqref{Higbm} of the second representation \eqref{first}, since $2\mu=\lambda$ and
\[s\frac{\Gamma(\mu-\kappa+1/2)}{\mu+\kappa-1/2}=2\Gamma\left(\frac12(\nu+\lambda)+1\right).\]

\er
We are also interested in the ``increments" of hitting times. For example, the time it takes for Yor's process to reach level $y$ after attaining level $x$,
\[H_y-H_x=\inf\{t>0: A^{(\nu)}_{H_x+t}=y\}.\]

\bp \label{prop:increment}
For $y>x>0$,
\[H_y-H_x \thicksim  H_{(R^{(\nu)}_x)^{-2}(y-x)},\] where $\{R^{(\nu)}_x, x \ge 0\}$ is a Bessel process with index $\nu$ independent of $A$ on the right-hand side.
\ep
\begin{proof} Note that
\by
\Pro(H_y-H_x \le t)&=&\Pro(A^{(\nu)}_{H_x+t} \ge y)\\
&=&\Pro\left(\int^{H_x}_0 \exp\{2B^{(\nu)}_s\}\df s+\int^{H_x+t}_{H_x} \exp\{2 B^{(\nu)}_s\} \df s \ge y \right)\\
%&=&\Pro\left( A^{(\nu)}_{H_x}+ \int^{H_x+t}_{H_x} \exp\{2 B^{(\nu)}_s\} \df s \ge y  \right)\\
&=&\Pro\left( \int^{H_x+t}_{H_x} \exp\{2 B^{(\nu)}_s\} \df s \ge y-x  \right)\\
&=&\Pro\left( \exp\{2B^{(\nu)}_{H_x}\}\int^{H_x+t}_{H_x} \exp\{2 (B^{(\nu)}_s-B^{(\nu)}_{H_x})\} \df s \ge y-x  \right).
\ey
Letting $s \rightarrow 0$ in \eqref{lap} yields that $\Pro(H_x<\infty)=1$ for all $x \ge 0$. Thus it follows immediately from Theorem 6.16 of \cite[page 86]{KarShr} that $\{W^{(\nu)}_t:=B^{(\nu)}_{H_x+t}-B^{(\nu)}_{H_x},t \ge 0  \}$ is a drifted Brownian motion independent of $\mathcal{F}_{H_x}$. Therefore,
\by
%&&\Pro\left( \exp\{2B^{(\nu)}_{H_x}\}\int^{H_x+t}_{H_x} \exp\{2 (B^{(\nu)}_s-B^{(\nu)}_{H_x})\} \df s \ge y-x  \right)\\
\Pro(H_y-H_x\le t)
%&=&\Pro\left( \exp\{2B^{(\nu)}_{H_x}\}\int^{t}_{0} \exp\{2 (B^{(\nu)}_{H_x+u}-B^{(\nu)}_{H_x})\} \df u \ge y-x  \right)\\
&=&\Pro\left( \exp\{2B^{(\nu)}_{H_x}\}\int^{t}_{0} \exp\{2 W^{(\nu)}_u\} \df u \ge y-x  \right)\\
&=&\Pro\left( \exp\{2B^{(\nu)}_{H_x}\} \ol{A}^{(\nu)}_t \ge y-x  \right),
\ey where $\ol{A}_t$ is a Yor's process independent of $B^{(\nu)}_{H_x}.$
Recall Lamperti's identity (c.f. \cite[(2.a)]{Yor92})
\[\exp\{B^{(\nu)}_t\}=R^{(\nu)}_{A^{(\nu)}_t},\qquad t \ge 0,\] where $\{R^{(\nu)}_t, t \ge 0\}$ is a Bessel process with index $\nu$ (starting from $1$). Therefore, it follows immediately that
\[\Pro(H_y-H_x \le t)=\Pro((R^{(\nu)}_x)^2 \ol{A}^{(\nu)}_t \ge y-x)=\Pro(\ol{A}^{(\nu)}_t \ge (R^{(\nu)}_x)^{-2} (y-x) )=\Pro(H_{(R^{(\nu)}_x)^{-2} (y-x)}\le t).\] Note that in this case $R^{(\nu)}_x$ is independent of $\ol{A}^{(\nu)}_t$.
\end{proof}

\bc The Laplace transform of $H_y-H_x$ for $y>x>0$ is given by
\be
\Exp[e^{-s(H_y-H_x)}]&=&\int^\infty_0 (2(y-x))^{(\nu-\lambda)/2}\exp\left\{-\frac{y\rho^2+y-x}{2x(y-x)}\right\}\frac{\Gamma((\nu+\lambda)/2+1)}{\Gamma(\lambda+1)}\nb \\
&& \times M\left(\frac12(\nu+\lambda)+1,\lambda+1;\frac{\rho^2}{2(y-x)} \right) \frac{\rho^{\lambda+1}}{x}I_\nu\left( \frac\rho x \right) \df \rho. \label{tauyxlap}
\ee
\ec
\begin{proof} It follows from Proposition \ref{prop:increment} that
\by \Exp[e^{-s(H_y-H_x)}]&=&\int^\infty_0 \Exp[e^{-s H_{\rho^{-2}(y-x)}}] \frac{\rho^{\nu+1}}{x}\exp \left( -\frac1{2x}(1+\rho^2) \right)I_\nu\left( \frac\rho x \right) \df \rho\\
&=& \int^\infty_0 (2\rho^{-2}(y-x))^{(\nu-\lambda)/2}\exp\{-\frac{\rho^2}{2(y-x)}\}\frac{\Gamma((\nu+\lambda)/2+1)}{\Gamma(\lambda+1)}\\
&& \times M\left(\frac12(\nu+\lambda)+1,\lambda+1;\frac{\rho^2}{2(y-x)} \right) \frac{\rho^{\nu+1}}{x}\exp \left( -\frac1{2x}(1+\rho^2) \right)I_\nu\left( \frac\rho x \right) \df \rho,
\ey which leads to \eqref{tauyxlap} with slight rearrangement.
\end{proof}

\section{Hitting time of diffusion process} \label{diffusion}

In the theory of interest, annuity-certain is a type of financial arrangement in which payments of fixed amount are made periodically. If the investor starts with $x$ dollars and makes continuous payments into a savings account at the rate of one dollar per time unit and the accounts earns interest at the rate of $\mu$ per time unit, then the accumulated value at time $t$ of the account is
\be \ol{s}_{\bx t}:=e^{\mu t}x+\int^t_0 e^{\mu (t-s) } \df s =e^{\mu t}x+\frac{e^{\mu t}-1}{\mu},\label{detcase}\ee or equivalently by the ODE
\[\df \ol{s}_{\bx t}=[1+\mu \ol{s}_{\bx t}\,] \df t,\qquad \ol{s}(0)=x.\] In generalization, if the accumulation of deposits is linked to an equity index driven by a geometric Brownian motion \[\df S_t=\mu S_t \df t +\sigma S_t \df B(t).\] Then the accumulated value at time $t$ of one dollar deposited at $u$ should be proportional to the financial return of buying one share of the equity index over the period $[u,t]$, i.e. $S_t/S_u.$ With analogy to \eqref{detcase}, the accumulated value at $t$ of incoming annuity payments should be
\be X(t)=x \frac{S_t}{S_0}+\int^t_0 \frac{S_t}{S_u} \df u.\label{Igen}\ee Similarly, with an initial deposit of $y$ dollars at time $0$, the outstanding balance at time $t$ of outgoing annuity payments would be
\be Y(t)=y \frac{S_t}{S_0}-\int^t_0 \frac{S_t}{S_u} \df u.\label{Dgen}\ee
A simple application of It\^o's formula shows that they are in fact the geometric Brownian motions with affine drift. We may write $X^{(\mu,\sigma)}$ and $Y^{(\mu,\sigma)}$ to indicate their dependency on parameters.
\by \df X_t&=&[\mu X_t+1 ] \df t+ \sigma \df B_t,\quad X_0=x.\\
\df Y_t&=& [\mu Y_t-1] \df t +\sigma \df B_t,\qquad Y_0=y.
\ey
One can also introduce annuity payments at any arbitrary constant number ($\pm 1$ replaced by any constant). However, the SDEs of such form achieve no more generality than \eqref{Isde} and \eqref{Dsde}, as one can convert one to another by changing the time scale. The relations \eqref{IA} and \eqref{tauDA} continue to hold except for a change in time parameter. For example, for $x=0$ and each fixed $t\ge 0$,
\be X^{(\mu,\sigma)}_t &\sim& \frac4{\sigma^2} A^{(\nu)}_{\sigma^2 t/4}\, ,\qquad \mbox{ with } \nu=\frac{2\mu-\sigma^2}{\sigma^2},\nb\\
\tau^{(\mu,\sigma)}_{y,0} &\sim& \frac{4}{\sigma^2} H^{(\nu)}_{\sigma^2 y/4},\qquad \mbox{ with } \nu=\frac{\sigma^2 - 2\mu}{\sigma^2}.\label{tauDAgen}\ee
We can use the connections of these hitting times to find analytical solutions to their densities.

%Now let us investigate another diffusion process that relates to $\ol{s}(t)$ defined by the SDE
%\[ \df D(t)=[-1+ \mu D(t) ] \df t+\sigma D(t) \df B(t),\quad D(0)=d.\] It is easy to show by Ito-Doeblin formula that $D(t)$ is given by
%\[ D(t)=\mathcal{E}_t \left(d- \int^t_0 \mathcal{E}^{-1}_s \df s\right).\] Define the first passage time of this diffusion process
%\[ H_a(d) :=\inf\{t: D_t<a, D_0=d\},\quad a \le d.\] Clearly, we have the relationship that
%\[ H_0(d) = \inf \left\{t: \int^t_0 \exp \{-(\mu-\frac{1}{2}\sigma^2)s-\sigma B_s\} \df s >d\right\}.\] Therefore, we can expect the duality of the two hitting times for any $d>0$,
%\[ H_0(d) \text{ has the same law as } C^{(\nu)}_d, \text{ when } \sigma=-2, \mu=2(1-\nu).\]

\bp \label{prop:tauD1}
The Laplace transform of $\tau^{(\mu,\sigma)}_{y,0}$ for $y>0$ is given by
\be \Exp^y[e^{-s \tau_{y,0}}]=\frac{\Gamma(b-k)}{\Gamma(b)}\left(\frac{2}{\sigma^2y} \right)^k \exp \left(-\frac{2}{\sigma^2 y} \right) M\left(b-k, b, \frac{2}{\sigma^2 y}  \right),\label{L} \ee where
\by
k:=\frac{2\mu-\sigma^2+ \sqrt{(2\mu-\sigma^2)^2+8\sigma^2 s}}{2\sigma^2},\quad b:=2k+2-\frac{2\mu}{\sigma^2}.
\ey
\ep

\begin{proof}
Denote the Laplace transform by $L_s(y)= \Exp [\exp \{-s \tau_{y,0}\}].$
According to Proposition 50.3 of \cite{RogWil} the solution is given by
\[L_s(y)=\frac{\phi_s (y)}{\phi_s(0)},\] where $\phi_s$ is a decreasing solution to the ODE
\be \frac{1}{2}\sigma^2 x^2 L''(x)+(-1+\mu x) L'(x)-s L(x)=0,\qquad x >0.\label{ODE}\ee We obtain two real-valued fundamental solutions to 
\[L_1(x)= x^{-k}\exp\left(-\frac{2}{\sigma^2x}\right)M(b-k,b,\frac{2}{\sigma^2x})=x^{-k}M(k,b,-\frac{2}{\sigma^2x}) \]
and
\[ L_2(x)=x^{-k}\exp\left(-\frac{2}{\sigma^2x}\right)U(b-k,b,\frac{2}{\sigma^2x}).\]
 where $M$ and $U$ are Kummer's functions of first and second kind respectively.
 It can be verified that $L_1$ is a decreasing solution whereas $L_2$ is an increasing solution. According to \cite{And}, we have the following asymptotics \[M(a,b,z)\sim \frac{\Gamma(b)}{\Gamma(a)} z^{a-b} e^z,\qquad z \rightarrow \infty.\]
Hence, as $x \rightarrow 0+$,
\[x^{-k}M\left(b-k,b, \frac{2}{\sigma^2 x}\right) \exp \left(-\frac{2}{\sigma^2 x} \right)\to \frac{\Gamma(b)}{\Gamma(b-k)}\left( \frac{2}{\sigma^2} \right)^{-k}.\]
Therefore, the solution to $L_s(y)$ is given by \eqref{L}. \end{proof}

\br \label{rem}
When matching the parameters by letting $\sigma=2, \mu=2(1-\nu),$ then $k=(\lambda-\nu)/2, b=\lambda+1$ and $b-k=(\nu+\lambda)/2+1$. It follows immediately that the two expressions \eqref{L} and \eqref{Higbm} agree, which confirms the identity in distribution of two hitting times \eqref{tauDAgen}. One should note, however, the result in Proposition \ref{prop:tauD1} is more general than that in Proposition \ref{prop:LapHa}, as we do not require $\mu\ge \sigma^2/2$, the equivalent of $\nu\ge 0$.
\er

\br \label{rem:connection}
One immediate use of the identity in distribution is that we could provide an explicit expression of the density function of $H_0(y)$, which is usually a very difficult task by other means. Since $(\sigma^2/4)\tau_0 \sim H_a,$ with $y=4a/\sigma^2$. Thus, if $\mu>\sigma^2/2$, we obtain the density of the hitting time $\tau_0$ given by
\[h(u)=\frac{\sigma^2}{4}f\left(\frac{\sigma^2}{4}u\right), \quad \text{ where } a=\sigma^2 d/4, v=1-2\mu/\sigma^2. \]
\er

For the convenience of applications in Section \ref{sec:policyholder}, we present here the density of $\tau_{y,0}$, even though similar results can be easily obtained for $\tau^{(\mu,\sigma)}_{y,0}$ as shown in Remark \ref{rem:connection}.

\bp
For $\nu\in \R$, the density of $\tau_{y,0}$ is given by
\be &&f(t)=\frac1{4\pi^2}(2y)^{(1-\nu)/2}\exp\left(-\frac1{4y} \right)\nb\\
&&\times\int^\infty_0 \frac{\nu^2+p^2}{2}\exp\left(-\frac{(\nu^2+p^2)t}{2} \right) W_{-(1-\nu)/2,ip/2}\left( \frac1{2y}\right) \left|\Gamma\left(\frac{-\nu+ip}{2}\right) \right|^2\sinh(\pi p) p \df p\nb\\
&&-I_{\{\nu> 2\}}\sum^{[\nu/2]}_{n=1}e^{-2n(\nu-n)t}\frac{2(-1)^n }{\Gamma(n)} \frac{(2y)^{-\nu+n}}{\Gamma(\nu-2n)}\;M\left(\nu-n,\nu-2n+1,-\frac1{2y}\right) .\qquad \label{denf}\ee The probability of eventual passage below zero is given by
\be
\Pro(\tau_{y,0}<\infty)=\left\{\begin{array}{ll} \frac{\gamma(\nu,1/(2y))}{\Gamma(\nu)}, & \qquad \nu>0;\\1 & \qquad \nu \le 0, \end{array} \right. \label{ultpass}
\ee where $\gamma$ is the lower incomplete Gamma function.
\ep

\begin{proof}
It is known in ~\cite[p863,(63)]{Lin04b} that for $\nu\in \R$, the density of $A^{(\nu)}_t$ is given by
\by
p(y)=\frac1{2\pi^2}\int^\infty_0 \exp\left\{-\frac{(\nu^2+p^2)t}{2}\right\}(2y)^{(\nu-1)/2}\exp(-\frac1{4y})W_{(1-\nu)/2,ip/2}\left(\frac1{2y}\right)\left| \Gamma\left(\frac{\nu+ip}{2}\right)\right|^2 \sinh(\pi p) p \df p\\
+I(\nu<0)\sum^{[|\nu|/2]}_{n=0}\exp(-2n(|\nu|-n)t)\frac{(-1)^n 2(|\nu|-2n)}{\Gamma(1+|\nu|-n)}(2y)^{n-1-|\nu|}\exp(-\frac1{2y})L^{(|\nu|-2n)}_n \left( \frac1{2y}\right).
\ey
Let us now consider the distribution function $\Pro(A^{(\nu)}_t<y)=\int^y_0 p(w) \df w.$ The integral part can be integrated using the method in ~\cite{FenVol}. We shall use the following identity derived from ~\cite[p.463, (2.19.3.5)]{PruBryMar} for the summation part. For $x>0, \Re(\mu+n)>0, n \in \Z^+,$
\[\int^\infty_x z^{\mu+n-1} e^{-z} L^{(\mu)}_n(z) \df z=-\frac{(\mu+1)_n}{n!}\frac{x^\alpha}{\alpha}\; _1F_1[ \mu+n;\nu+1;-x].\]
Making a simple change of variables gives
\by&&\int^y_0 (2w)^{-\mu-n-1}\exp(-\frac1{2w})L^{(\mu)}_n(\frac1{2w}) \df w=-\frac{(\mu+1)_n}{2(n!)} \frac{(2y)^{-(\mu+n)}}{\mu+n}\;_1F_1\left[\mu+n;\mu+1;-\frac1{2y}\right].\ey
Note that the first term in the summation has a simpler form
\[\frac2{\Gamma(|\nu|)}(2y)^{-1-|\nu|}\exp(-\frac1{2y}).\]
Thus, using a change of variables yields
\[\frac2{\Gamma(|\nu|)}\int^y_0 (2z)^{-1-|\nu|}\exp(-\frac1{2z}) \df z=\frac{\Gamma(|\nu|,1/(2y))}{\Gamma(|\nu|)}.\]
In summary, the distribution of $A^{(\nu)}_t$ is given by
\be
&&\Pro(A^{(\nu)}_t<y)=\frac1{4\pi^2}(2y)^{(1+\nu)/2}\exp\left(-\frac1{4y} \right)\nb\\
&&\times\int^\infty_0 \exp\left(-\frac{(\nu^2+p^2)t}{2} \right) W_{-(1+\nu)/2,ip/2}\left( \frac1{2y}\right) \left|\Gamma\left(\frac{\nu+ip}{2}\right) \right|^2\sinh(\pi p) p \df p\nb\\
&&+I(\nu<0) \frac{\Gamma(|\nu|,1/(2y))}{\Gamma(|\nu|)} -I(\nu< -2)\sum^{[|\nu|/2]}_{n=1}\exp(-2n(|\nu|-n)t)\frac{(-1)^n (|\nu|-2n)}{\Gamma(1+|\nu|-n)} \nb \\
&&\times \left\{\frac{(|\nu|-2n+1)_n}{n!}\frac{(2y)^{-|\nu|+n}}{|\nu|-n}M\left(|\nu|-n,|\nu|-2n+1,-\frac1{2y}\right)  \right\}, \qquad \label{distA}\ee where we used the fact that Kummer function $M$ is the same as the hypergeometric function $_1F_1$.
%It was shown in the previous version of Feng and Volkmer~\cite{FenVol2} that for $\nu \ge 0$
%\[\Exp\left[ e^{ -s H_a }  \right]=(2a)^{(\nu-\lambda)/2} \exp \left(-\frac{1}{2a}  %\right)\frac{\Gamma((\nu+\lambda)/2+1)}{\Gamma(\lambda+1)} M\left(\frac{1}{2}(\nu+\lambda)+1, %\lambda+1; \frac{1}{2a}\right),\] where $\lambda=\sqrt{2s+\nu^2}$ and $H_a:=\inf\{t: A^{(\nu)}_t=a\}.$ %Although the proof for Proposition 2.2 does not work for $\nu<0$, the expression works for all $\nu\in \R$ %due to Remark 3.1.
Since $\Pro(\tau_{y,0}>t)=\Pro(H^{(-\nu)}_y>t)=\Pro(A^{(-\nu)}_t<y),$ we differentiate the above expression w.r.t. $t$ and obtain the density function \eqref{denf}.

It follows from Proposition \ref{prop:tauD1} and Remark  \ref{rem} that for $s>0$
\be \Exp\left[ e^{ -s \tau_{y,0} }  \right]=(2y)^{(-\nu-\lambda)/2} \exp \left(-\frac{1}{2y}  \right)\frac{\Gamma((-\nu+\lambda)/2+1)}{\Gamma(\lambda+1)} M\left(\frac{1}{2}(-\nu+\lambda)+1, \lambda+1; \frac{1}{2y}\right),\label{Laptau}\ee where $\lambda=\sqrt{2s+\nu^2}$. 
When $s=0,\nu\le 0$, $\lambda=|\nu|=-\nu$, we obtain
\by \Pro(\tau_{y,0}<\infty)=\exp\left( -\frac1{2y}\right)M\left(1-\nu,1-\nu,\frac1{2y}\right)=1.\ey When $\nu>0, \lambda=\nu,$ since \[\Pro(\tau_{y,0}=\infty)=\lim_{t\to \infty} \Pro(A^{(-\nu)}_t<y)=\frac{\Gamma(\nu,1/(2y))}{\Gamma(\nu)},\] we must have, for $\nu>0$,
\by \Pro(\tau_{y,0}<\infty)=\frac{\gamma(\nu,1/(2y))}{\Gamma(\nu)}<1.\ey
\end{proof}

We can also explore a ``symmetry" between the processes $X$ and $Y$. It is clear from \eqref{Igen} and \eqref{Dgen} that for any $X(0)=-Y(0) \in \R$,
\[\{X(t), t \ge 0\} \mbox{ has the same law as } \{-Y(t), t \ge 0\}.\] Introduce the hitting time of $X$
\[\rho^{(\mu,\sigma)}_{x,y}:=\inf\{t: X_0 =x, X_t=y\}.\] Then it follows that for any $X(0)=-Y(0)=x \in \R,$ 
\be \rho^{(\mu,\sigma)}_{x,y} \sim \tau^{(\mu,\sigma)}_{-x,-y},\qquad y\in \R.\label{tauID}\ee

\bp
The Laplace transform of $\tau^{(\mu,\sigma)}_{x,y}$ is given by
\[ \Exp\left[ e^{-s \tau^{(\mu,\sigma)}_{x,y}}  \right]=\left\{ \begin{array}{ll}
\displaystyle \left( \frac{x}{y}\right)^{-k} \frac{M(k,b,-2/(\sigma^2 x))}{M(k,b,-2/(\sigma^2 y))} ,&\qquad x>y>0 \mbox{ or } 0>y>x,\\
\displaystyle \left( \frac{x}{y}\right)^{-k}\exp\left(-\frac{2}{\sigma^2 x}+\frac{2}{\sigma^2 y}  \right) \frac{U(k,b,2/(\sigma^2 x))}{U(k,b,2/(\sigma^2 y))} ,&\qquad y>x>0,\\
\displaystyle \left( \frac{x}{y}\right)^{-k}\frac{U(k,b,-2/(\sigma^2 x))}{U(k,b,-2/(\sigma^2 y))},&\qquad 0>x>y. \end{array} \right.\]
\ep
\begin{proof}
We first consider $x>y>0$. It is clear from the proof of Proposition \ref{prop:tauD1} that as $L_1, L_2$ are the decreasing and increasing solutions respectively for $x>0$. Thus, 
\by \Exp\left[ e^{-s \tau^{(\mu,\sigma)}_{x,y}}  \right]=\frac{L_1(x)}{L_1(y)},\qquad \Exp\left[ e^{-s \tau^{(\mu,\sigma)}_{y,x}}  \right]=\frac{L_2(y)}{L_2(x)},\ey the first of which yields the expression for $x>y>0$ and the second of which produces the expression for $y>x>0$ by exchanging $x$ and $y$. When $x<0$, we obtain two real-valued fundamental solutions to \eqref{ODE}
\by
L_1(x)=x^{-k}M(k,b,-\frac{2}{\sigma^2 x}),\\
L_2(x)=x^{-k}U(k,b,-\frac{2}{\sigma^2 x}).
\ey In this case, $L_2$ is the decreasing function whereas $L_1$ is the increasing function. Thus for $y<x<0$, 
\by \Exp\left[ e^{-s \tau^{(\mu,\sigma)}_{x,y}}  \right]=\frac{L_2(x)}{L_2(y)},\ey and for $0>x>y$  and $0>y>x$,\by \Exp\left[ e^{-s \tau^{(\mu,\sigma)}_{y,x}}  \right]=\frac{L_1(y)}{L_1(x)},\ey which yield the desired expressions.
\end{proof}

\bc
The Laplace transform of $\rho^{(\mu,\sigma)}_{x,y}$ is given by
\[ \Exp\left[ e^{-s \rho^{(\mu,\sigma)}_{x,y}}  \right]=\left\{ \begin{array}{ll}
\displaystyle \left( \frac{x}{y}\right)^{-k} \frac{M(k,b,2/(\sigma^2 x))}{M(k,b,2/(\sigma^2 y))} ,&\qquad x>y>0 \mbox{ or } 0>y>x,\\
\displaystyle \left( \frac{x}{y}\right)^{-k}\frac{U(k,b,2/(\sigma^2 x))}{U(k,b,2/(\sigma^2 y))},&\qquad y>x>0, \\
\displaystyle \left( \frac{x}{y}\right)^{-k} \exp\left(\frac2{\sigma^2 x}-\frac2{\sigma^2 y} \right)\frac{U(k,b,-2/(\sigma^2 x))}{U(k,b,-2/(\sigma^2 y))} ,&\qquad 0>x>y. \end{array} \right.\]
\ec
\begin{proof} Use the identity in distribution \eqref{tauID} to obtain
\[\Exp\left[ e^{-s \rho^{(\mu,\sigma)}_{x,y}}  \right]=\Exp\left[ e^{-s \tau^{(\mu,\sigma)}_{-x,-y}}  \right],\] which yields the desired expressions.
\end{proof}

\br
Passing to the limit as $x \rightarrow 0$, we obtain for $y>0$,
\[\Exp[e^{-s \rho^{(\mu,\sigma)}_{0,y}}]=\left( \frac2{\sigma^2 y} \right)^{-k} \frac1{U(k,b,2/(\sigma^2 y))}.\]
\er

%{\bf (The following part may be removed)}

%In the case of annuities-certain, we know exactly at what time the accumulated value hits a certain target %level. For example, if $\ol{s}(t)=0, \ol{s}(t)=(e^{\mu t}-1)/\mu.$ Then the hitting time $\tau_y:=\inf\{t: %\ol{s}(t)\ge y\}$ for $y>0$ is deterministic, i.e.
%\[\tau_y=\frac{\ln(1+\mu y)}{\mu}.\] Then the Laplace transform of such a degenerate random variable %must be
%\[\Exp^0[e^{-s \tau_y}]=(1+\mu y)^{-s/\mu}.\]

%\bp
%For $y>0$,
%\[E^0\big[ e^{-s \tau_y(X)}  \big]\rightarrow (1+\mu y)^{-s/\mu}, \qquad \mbox{ as } \sigma \rightarrow %0.\]
%\ep
%\begin{proof}
%Let $z=2/(\sigma^2 y)$. It is easy to show that $k \rightarrow -s/\mu$ and $b/z \rightarrow -\mu,$ as $\sigma \rightarrow 0.$ Note that the integral representation for $U$ is
%\[U(a,b,z)=\frac1{\Gamma(a)}\int^\infty_0 e^{-x t} t^{a-1}(1+t)^{b-a-1} \df t,\qquad a , x >0.\]
%\end{proof}
%\bp
%For $y>0$,
%\[E^y\big[ e^{-s \tau_0(Y)} \big]=\left\{ \begin{array}{ll} (1-\mu y)^{s/\mu}, & \qquad x<\frac{1}{\mu};\\ %0,&\qquad x \ge \frac{ 1}{\mu}, \end{array}  \right. \qquad \mbox{ as } \sigma \rightarrow 0.\]
%\ep

\section{Guaranteed Minimum Withdrawal Benefit}

The variable annuity guarantee product is arguably the most complex investment-combined insurance policy available to individual investors. Without any investment guarantees, they are almost the same as mutual funds except that all purchase payments are tax-deferred. In order to compete with mutual funds, variable annuity writers have introduced a variety of investment guarantees, among which the most recent market innovation is the guaranteed minimum withdrawal benefit (GMWB). \cite{MilSal} was among the first to provide a mathematical model for the valuation of the GMWB, followed by various optimization problems based on withdrawal strategies in \cite{DaiKwoZon}, \cite{CheFor}, \cite{ForVet}, all of which are based on numerical PDE solutions. In this work, we attempt to show that in the plain vanilla case, the fair charge of the GMWB rider can in fact be determined by analytical solutions.

The GMWB typically is sold as a rider to variable annuity contracts. The policyholder is allowed to withdraw up to a fixed amount per year out of the investment fund without penalty. On the liability side, the GMWB rider guarantees the return of total purchase payment regardless of the performance of the underlying investment funds.  For example, a contract starts with an initial purchase payment of \$100 and the policyholder elects to withdraw the maximum amount $7\%$ of purchase payment without penalty each year. Due to the poor performance of the funds in which the policyholder invests, the account value is depleted at the end of five years. By this time, the policyholder would have only withdrawn $\$7 \times 5 =\$ 35$ in total. Then the guarantee kicks in to sustain the annual withdrawal  $\$7$ until the entire purchase payment is return, which means it pays until  the maturity at the end of $\$100/\$7=14.28$ years.
On the revenue side, the guarantee is funded by daily charges of a fixed percentage of fees from the investment funds.

\subsection{Policyholder's perspective}
\label{sec:policyholder}

Viewed as an investment vehicle, the GMWB rider is priced under the no-arbitrage assumptions from the investor's perspective in \cite{MilSal}. Let $G$ be the initial deposit and $w$ be the guaranteed rate of withdrawal per time unit. Thus the GMWB rider provides safeguards to the continuous withdrawal until the initial deposit is completely refunded, i.e. the GMWB matures at time $T=G/w$.
Let $r$ be the risk-free force of interest. Thus the present value of guaranteed income is
\[w\int^T_0 e^{-rs} \df s=\frac{w}{r}(1-e^{-rT}).\]  In addition, if the equity fund performed well and the fund is not exhausted at maturity, then the policyholder is entitled to the then-current fund balance. We assume that the equity-index is driven under the physical measure by a geometric Brownian motion
\[\df S_t= \mu^\ast S_t \df t+\sigma S_t \df \td{W}_t.\] As with mutual funds, the policyholder's investment fund is linked to the equity index so that its value fluctuates in proportion to the equity index.
 Let $m>0$ be the rate per time unit of total fees charged by the insurer as a fixed percentage of the fund. Then the fund value is driven by
\[\df F_t=[(\mu^\ast-m) F_t - w] \df t +\sigma F_t \df \td{W}_t, \qquad F_0=G>0.\]
Throughout the section, we shall denote $\tau_0(F):=\inf\{t: F_t=0\}$ and reserve $\tau_0:=\tau_{y,0}$ for the process $Y$ whenever $Y(0)=y$ is clear from the context. Thus the policyholder receives at maturity
\[F_T I(\tau_0(F) >T),\] We determine the fair value of fees $m$ so that the arbitrage-free price of policyholder's asset at maturity plus guaranteed income is equal to the initial purchase payment, i.e. to find $m$ such that
\be\Exp_Q[e^{-rT}F_TI(\tau_0>T)]+\frac{w}{r}(1-e^{-rT})=G,\label{pol}\ee where $\Exp_Q$ is the risk-neutral measure. 
Under the risk neutral measure, the dynamics of the investment fund is driven by
\[\df F_t=[(r-m) F_t - w] \df t +\sigma F_t \df W_t, \qquad F_0=G>0,\] where $W$ is the corresponding Brownian motion under the risk neutral measure.
Note that $\sigma W(t)\sim 2 B_{\sigma^2 t/4},$ where $B$ is an independent Brownian motion. We let
\[t:=\frac{\sigma^2 T}{4},\qquad Y_t:=\frac{\sigma^2}{4w}F_T.\] Then the process $Y$ satisfies \eqref{Dsde} with $\nu:=[2(r-m)-\sigma^2]/\sigma^2.$ Thus $\tau_0(F)=(4/\sigma^2)\tau_0$ and 
\[\Exp_Q[F_TI(\tau_0(F)>T)]=\frac{4w}{\sigma^2}h(t), \] where \be h(t):=\Exp^y[ Y_t I(\tau_0>t)],\qquad y:=\frac{\sigma^2 G}{4w}.\label{defh}\ee Here we use $\Exp^y$ to indicate the risk-neutral measure under which $Y(0)=y$.

\bp \label{prop:h} Let $\lambda:=|\nu+2|, \kappa:=(1-\nu)/2.$ If $\nu\ne -1$, then
\be
h(t)&=&\left(y-\frac1{2(\nu+1)}\right)e^{2(\nu+1)t}+I_{\{\nu>0\}} \frac{\Gamma(\nu,1/(2y))}{2(\nu+1)\Gamma(\nu)}\nb\\
&&+\frac{e^{2(\nu+1)t}}{2(\nu+1)}(2y)^{(-\nu-\lambda)/2} \exp \left(-\frac{1}{2y}  \right)\frac{\Gamma(\frac{-\nu+\lambda}{2}+1)}{\Gamma(\lambda+1)} M\left(\frac{-\nu+\lambda}{2}+1, \lambda+1; \frac{1}{2y}\right)\nb\\
&&+\frac{(2y)^\kappa}{8\pi^2} \exp\left(-\frac{1}{4y}\right) 
\int_0^\infty e^{-(\nu^2+p^2)t/2} W_{-\kappa, ip/2}\left(\frac1{2y}\right) 
\left|\Gamma\left(\frac{ip}{2}-\frac\nu2 -1\right)\right|^2 \sinh(\pi p) p\,dp\nb\\
&&-I_{\{\nu>2\}}\sum^{[\nu/2]}_{n=1}\frac{(-1)^n (2y)^{-\nu+n}e^{-2n(\nu-n)t}}{(\nu-n)[2(\nu+1)+2n(\nu-n)]\Gamma(n+1)\Gamma(\nu-2n)}M\left(\nu-n;\nu-2n+1,-\frac1{2y}\right).\qquad \label{h}
\ee
If $\nu=-1$, then
\be
h(t)=ye^{-\frac{1}{2y}}-\frac12 E_1\left(\frac1{2y}\right)+\frac{y}{4\pi^2} e^{-\frac1{4y}}\int^\infty_0 e^{-\frac{(1+p^2)t}2}W_{-1,ip/2}\left( \frac1{2y}\right)\left|\Gamma\left(\frac{-1+ip}2\right) \right|^2\sinh(\pi p) p \df p,\qquad \label{hnu=-1}
\ee where $E_1$ is the exponential integral.
\ep

\begin{proof}
Recall that $\{\tau_0>t\}\equiv \{Y_t>0\}.$ Hence $h(t)=\Exp[Y_t I(Y_t>0)].$ Thus, it suffices to find an expression for $\Exp[Y_tI(Y_t< 0)]. $ Observe that using the strong Markov property
\by
\Exp^y[Y_tI(Y_t<0)]&=&\Exp^y\Big[\left.\Exp[Y_t I(\tau_0<t)\right|\mathcal{F}_{\tau_0}] \Big]=\Exp^y\Big[ I(\tau_0<t)\Exp^0[Y_{t-\tau_0}]\Big].
\ey
First consider $\nu\ne -1$. It is easy to show that the mean of the process $Y$ is
\[\Exp^y[Y_t]=y e^{2(\nu+1)t}-\frac1{2(\nu+1)}\left(e^{2(\nu+1)t}-1\right).\] 
Thus, we want to compute the integral
\[ g(t):=\Exp^y[Y_tI(Y_t<0)]=\frac{1}{2(\nu+1)} \int_0^t (1-e^{2(\nu+1)(t-u)})f(u)\,du .\]
Once $g(t)$ is known, we can obtain $h(t)=\Exp[Y_t]-g(t).$
To simplify the double integral in $g(t)$, we rewrite $g(t)$ in the form 
\[ g(t)=\frac{1}{2(\nu+1)}\int^\infty_0 f(u) \df u -\frac{e^{2(\nu+1)t}}{2(\nu+1)} \int_0^\infty e^{-2(\nu+1)u} f(u)\,du -\frac{1}{2(\nu+1)} \int_t^\infty (1-e^{2(\nu+1)(t-u)})f(u)\,du. 
\] We now distinguish the cases $\nu \le 0$ and $\nu>0$.\\

\noindent {\bf When $ \nu\le 0$,}\\

The first integral is determined by the lower case of \eqref{ultpass}. We show in Lemma \ref{lem:ext} that the Laplace transform can be extended to $s=2(\nu+1)$ even if $\nu<-1$, as $2(\nu+1)\ge -\nu^2/2$. Thus,
\be \int_0^\infty e^{-2(\nu+1)u} f(u)\,du=(2y)^{(-\nu-\lambda)/2} \exp \left(-\frac{1}{2y}  \right)\frac{\Gamma((-\nu+\lambda)/2+1)}{\Gamma(\lambda+1)} M\left(\frac{1}{2}(-\nu+\lambda)+1, \lambda+1; \frac{1}{2y}\right), \qquad \label{lapext}\ee where $\lambda=\sqrt{4(\nu+1)+\nu^2}=|\nu+2|.$ It is also clear from the proof of Lemma \ref{lem:ext} that the integrand of $f$ is absolutely integrable over $[t,\infty)$ for $t>0$. We can exchange the order of integration for the third integral by Fubini's Theorem. Piecing all together, we get
\by
g(t)&=&\frac1{2(\nu+1)}-\frac{e^{2(\nu+1)t}}{2(\nu+1)}(2y)^{(-\nu-\lambda)/2} \exp \left(-\frac{1}{2y}  \right)\frac{\Gamma(\frac{-\nu+\lambda}{2}+1)}{\Gamma(\lambda+1)} M\left(\frac{-\nu+\lambda}{2}+1, \lambda+1; \frac{1}{2y}\right)\\
&&-\frac{(2y)^\kappa}{8\pi^2} \exp\left(-\frac{1}{4y}\right) 
\int_0^\infty e^{-(\nu^2+p^2)t/2} W_{-\kappa, ip/2}\left(\frac1{2y}\right) 
\left|\Gamma\left(\frac{ip}{2}-\frac\nu2 -1\right)\right|^2 \sinh(\pi p) p\,dp.
\ey

\noindent {\bf When $\nu>0$,}\\

Similarly, the first integral is determined by the top case of \eqref{ultpass}. The second integral is given by \eqref{lapext} due to \eqref{Laptau}. We use Fubini's theorem to obtain the expression for the third integral.
\by
g(t)&=&\frac{\gamma(\nu,1/(2y))}{2(\nu+1)\Gamma(\nu)}-\frac{e^{2(\nu+1)t}}{2(\nu+1)}(2y)^{(-\nu-\lambda)/2} \exp \left(-\frac{1}{2y}  \right)\frac{\Gamma(\frac{-\nu+\lambda}{2}+1)}{\Gamma(\lambda+1)} M\left(\frac{-\nu+\lambda}{2}+1, \lambda+1; \frac{1}{2y}\right)\\
&&-\frac{(2y)^\kappa}{8\pi^2} \exp\left(-\frac{1}{4y}\right) 
\int_0^\infty e^{-(\nu^2+p^2)t/2} W_{-\kappa, ip/2}\left(\frac1{2y}\right) 
\left|\Gamma\left(\frac{ip}{2}-\frac\nu2 -1\right)\right|^2 \sinh(\pi p) p\,dp\\
&&+I_{\{\nu>0\}}\sum^{[\nu/2]}_{n=1}\frac{(-1)^n (2y)^{-\nu+n}e^{-2n(\nu-n)t}}{(\nu-n)[2(\nu+1)+2n(\nu-n)]\Gamma(n+1)\Gamma(\nu-2n)}M\left(\nu-n;\nu-2n+1,-\frac1{2y}\right).
\ey

\noindent Then consider $\nu=-1$. Note that $\Exp^y(Y_t)=y-t$. Thus,
\[g(t)=-t \int^t_0 f(u) \df u +\int^\infty_0 u f(u) \df u-\int^\infty_t u f(u) \df u.\]
We denote the three terms by $g_1, g_2, g_3$ respectively. Note that
\by
g_1(t)=-t\Pro(\tau^{(-1)}_0<t)=-\frac{ty}{2\pi^2}e^{-\frac1{4y}}\int^\infty_0 e^{-\frac{(1+p^2)t}{2}}W_{-1,ip/2}\left( \frac1{2y}\right)\left| \Gamma\left(\frac{1+ip}2 \right)\right| \sinh(\pi p)p \df p,
\ey where we used $\Pro(\tau^{(-1)}_0<t)=\Pro^y(A^{(1)}_t>y)$ known from \eqref{distA}. 
\by
g_2(t)=\Exp(\tau^{(-1)}_0)=\left.\frac{\partial}{\partial s} \Exp(e^{-s\tau^{(-1)}_0})\right|_{s=0}.
\ey
We observe that when $\nu=-1$ the expression \eqref{Laptau} reduces to
\[\Exp(e^{-s\tau^{(-1)}_0})=\frac{\sqrt{2\pi}}{4 \sqrt{y}} e^{-\frac1{4y}}\left[ (2y\sqrt{1+2s} +2y+1) I_{\frac{\sqrt{1+2s}}{2}} \left( \frac1{4y} \right)+I_{\frac{\sqrt{1+2s}}2+1}\left( \frac1{4y}\right)\right].\]
Using the identities \cite[p.251, (10.29.1)]{NIST} $I_{\nu+1}(z)=I_{\nu-1}(z)-(2\nu/z)I_\nu(z)$ and \cite[p.254,(10.38.6)]{NIST}
\[\left. \frac{\partial I_\nu(z)}{\partial \nu}\right|_{\nu \pm \frac12}= -\frac1{\sqrt{2\pi x}} (E_1(2x)e^x \pm \mathrm{Ei}(3x) e^{-x}),\] where $\mathrm{Ei}$ is the generalized exponential integral, we can show that
\[\left.\frac{\partial}{\partial s} \Exp(e^{-s\tau^{(-1)}_0})\right|_{s=0}=2ye^{-\frac1{4y}}\sinh\left(\frac1{4y}\right)+\frac12 E_1\left( \frac1{2y}\right).\]
Using Fubini's theorem, we can show that
\[g_3(t)=\frac{y}{2\pi^2}e^{-\frac1{4y}}\int^\infty_0 \left( t+\frac2{1+p^2}\right)e^{-\frac{(1+p^2)t}{2}}W_{-1,ip/2}\left( \frac1{2y}\right)\left| \Gamma\left(\frac{1+ip}2 \right)\right| \sinh(\pi p)p \df p.\] Combining all terms we arrive at \eqref{hnu=-1} after simplifications.
\end{proof}

\br In the case that $\nu<-2$, $h(t)$ can be further simplified to
\be h(t)=\frac{(2y)^\kappa}{8\pi^2} \exp\left(-\frac{1}{4y}\right) 
\int_0^\infty e^{-(\nu^2+p^2)t/2} W_{-\kappa, ip/2}\left(\frac1{2y}\right) 
\left|\Gamma(i\frac{p}{2}-\frac\nu2 -1)\right|^2 \sinh(\pi p) p\,dp .\qquad \label{h2}\ee
This result can be obtained from the spectral method used in \cite{FenVol2} to determine risk measures (Section 3.2). 

We can verify that \eqref{h} agrees with \eqref{h2}. In the case where $\nu<-2$, we must have $\nu+\lambda=-2-2\nu$ and $\nu-\lambda=2$. Note that this expression agrees with the expression from the spectral method if 
\[y=\frac1{2(\nu+1)}\left[1-(2y)\exp(-\frac1{2y})\frac{\Gamma(-\nu)}{\Gamma(-\nu-1)}M\left(-\nu,-\nu-1,\frac1{2y}\right) \right].\] This is equivalent to
\[M\left(-\nu,-\nu-1,\frac1{2y}\right)=\frac{2(\nu+1)y-1}{2(\nu+1)y}\exp\left(\frac1{2y}\right),\] which can be easily proved using the series representation of $M$.
\er

\subsection{Insurer's perspective}

We can also price the GMWB from an insurer's point of view so that the insurer's revenue covers its liability. The outgoing cash flow for the insurer is the guaranteed payments after the investment fund is exhausted prematurely,
\[w\int^T_{\tau_0(F)} e^{-rs} \df s I(\tau_0(F) <T)=\frac{w}{r}(e^{-r\tau_0(F)}-e^{-rT})I(\tau_0(F)<T),\] and the incoming cash flow for the insurer is determined by the GMWB rider charges 
\[m_w\int^{\tau_0(F) \wedge T}_0 e^{-rs} F_s \df s,\] where $m_w$ is the rate per time unit of fees allocated to fund the GMWB. Note that in general $m> m_w$ as part of fees and charges are used to cover overheads and other expenses.  Assuming both cash flows can be securitized as tradable assets, we can also use the no-arbitrage arguments to determine the fair fees $m$ by
\be \frac{w}{r}\Exp_Q\left[e^{-r\tau_0(F)}I(\tau_0(F)<T)\right]-\frac{w}{r}e^{-rT} \Pro_Q\left(\tau_0(F)<T\right)=m_w\Exp_Q\left[ \int^{\tau_0(F) \wedge T}_0 e^{-rs} F_s \df s\right],\label{ins}\ee
Observe that the first term can be computed from
\[\Exp_Q\left[e^{-r\tau_0(F)}I(\tau_0(F)<T)\right]=\Exp^y\left[e^{-\hat{r}\tau_0}\right]-\Exp^y\left[e^{-\hat{r}\tau_0}I(\tau_0>t)\right]=:a(t,y)-b(t,y),\] where 
$ \hat{r}:=4r/\sigma^2,$ $a(t,y)$ is known from \eqref{Laptau}. The second term involves
\[c(t,y):=\Pro_Q(\tau_0(F)<T)=\Pro^y(\tau_0<t).\] The third term can be written as
\[\Exp_Q\left[ \int^{\tau_0(F) \wedge T}_0 e^{-rs} F_s \df s\right]=\frac{16w}{\sigma^4}d(t,y),\qquad d(t,y):=\Exp^y\left[ \int^{\tau_0\wedge t}_0 e^{-\hat{r}u} Y_u \df u  \right].\]
In the rest of this subsection, we derive explicit expressions for each of the three unknown quantities.

\bp
\by
b(t,y)=\frac1{\pi^2}(2y)^{\kappa}e^{-\frac{1}{4y}}\int^\infty_0 \frac1{\nu^2+p^2+2\hat{r}}e^{-\frac{(\nu^2+p^2+2\hat{r})t}{2}} W_{-\kappa,ip/2}\left( \frac1{2y}\right) \left|\Gamma\left(\frac{-\nu+ip}{2}+1\right) \right|^2 \sinh(\pi p)p \df p \\
-I_{\{\nu>2\}} \sum^{[\nu/2]}_{n=1} \frac{2(-1)^n(2y)^{-\nu+n}}{[2n(\nu-n)+\hat{r}]\Gamma(n)\Gamma(\nu-2n)}e^{-[2n(\nu-n)+\hat{r}]t}M\left(\nu-n,\nu-2n+1,-\frac1{2y}\right).\ey
\ep
\begin{proof} It follows that
$\Exp^y[e^{-\hat{r} \tau_0}I(\tau_0>t)]=\int^\infty_t e^{-\hat{r} u} f(u) \df u$ where $f$ is given in \eqref{denf}.
\end{proof}
\bp
\by &&c(t,y)=1-\frac1{4\pi^2}(2y)^{\kappa}e^{-\frac1{4y} }\int^\infty_0 e^{-\frac{(\nu^2+p^2)t}{2} } W_{-\kappa,ip/2}\left( \frac1{2y}\right) \left|\Gamma\left(\frac{-\nu+ip}{2}\right) \right|^2\sinh(\pi p) p \df p\\
&&-I_{\{\nu>0\}} \frac{\Gamma(\nu,1/(2y))}{\Gamma(\nu)} +I_{\{\nu>2\}}\sum^{[\nu/2]}_{n=1} e^{-2n(\nu-n)t}\frac{(-1)^n (\nu-2n)_{n}(2y)^{-\nu+n}}{\Gamma(1+\nu-n) n!}M\left(\nu-n,\nu-2n+1,-\frac1{2y}\right). \ey
\ep
\begin{proof}
It follows immediately from $\Pro(\tau_0<t)=\Pro^y(A^{(-\nu)}_t>y)$ which can be obtained from \eqref{distA}.
\end{proof}
\bp Let $\kappa=(1-\nu)/2$ and $\lambda=\sqrt{\nu^2+2\hat{r}}$.
If $\nu\ne -1$, then
 \be 
&&d(t,y)=y\left(\frac1{\hat{r}-2\nu-2}-\frac{e^{-(\hat{r}-2\nu-2)t}}{\hat{r}-2\nu-2}  \right)-\frac1{\hat{r}(\hat{r}-2\nu-2)}-\frac{e^{-\hat{r}t}}{\hat{r}(2\nu+2)}+\frac{e^{-(\hat{r}-2\nu-2)t}}{(\hat{r}-2\nu-2)(2\nu+2)}\nb\\
&&-\frac1{4\pi^2}(2y)^\kappa e^{-\frac1{4y} } \int^\infty_0 \frac1{2\hat{r}+\nu^2+p^2}e^{-\frac{(\nu^2+p^2+2\hat{r})t}2 } \sinh(\pi p) W_{-\kappa, ip/2}\left( \frac1{2y} \right) \left| \Gamma \left(\frac{ip}2-\frac{\nu}2-1\right) \right|^2 p \df p \nb\\
&&+\frac{\Gamma(1+\frac{\lambda-\nu}2)(2y)^{-\frac{\lambda+\nu}2}}{\hat{r}(\hat{r}-2\nu-2)\Gamma(1+\lambda)}M\left( \frac{\lambda+\nu}2,1+\lambda,-\frac1{2y} \right)+e^{-\hat{r}t}\frac{\Gamma(1+\frac{|\nu|-\nu}2)(2y)^{-\frac{|\nu|+\nu}2}}{\hat{r}(2\nu+2)\Gamma(1+|\nu|)} M\left( \frac{|\nu|+\nu}2,1+|\nu|,-\frac1{2y} \right)\nb\\
&&-e^{-(\hat{r}-2\nu-2)t}\frac{\Gamma(1+\frac{|\nu+2|-\nu}2)}{(\hat{r}-2\nu-2)(2\nu+2)\Gamma(1+|\nu+2|)(2y)^{-\frac{|\nu+2|+\nu}2}}M\left( \frac{|\nu+2|+\nu}2,1+|\nu+2|,-\frac1{2y} \right)\nb\\
&&-I_{\{\nu>2\}} \sum^{[\nu/2]}_{k=1} \frac{e^{-[\hat{r}+2k(\nu-k)]t} (-1)^{k}(2y)^{k-\nu}}{k(k-\nu)[2k(k-\nu)-\hat{r}][2k(k-\nu)-2\nu-2]\Gamma(k)\Gamma(\nu-2k)}M\left(\nu-k,\nu-2k+1,-\frac1{2y}  \right),\qquad \label{nune-1}
\ee 
If $\nu=-1$, then $\lambda=\sqrt{2\hat{r}+1}$ and
 \be 
&&d(t,y)=\frac{\hat{r}y-1}{\hat{r}^2}+\frac{\Gamma(1+\frac{\lambda+1}2)}{\hat{r}^2 \Gamma(1+\lambda)}(2y)^{-\frac{\lambda-1}2} M\left( \frac{\lambda-1}2,1+\lambda,-\frac1{2y} \right)-\frac{e^{-\hat{r}t}}{\hat{r}}\left[ye^{-\frac1{2y}}-\frac12 E_1\left(\frac1{2y}\right)\right]\nb\\
&&-\frac1{2\pi^2}y  e^{-\frac1{4y} } \int^\infty_0 \frac1{2\hat{r}+1+p^2}e^{-\frac{(1+p^2+2\hat{r})t}2 } \sinh(\pi p) W_{-1, ip/2}\left( \frac1{2y} \right) \left| \Gamma \left(\frac{ip}2-\frac12\right) \right|^2 p \df p. 
\qquad \label{nu=-1}
\ee
\ep

\begin{proof}
We take a Laplace transform to remove the finite time $t$. Define
\[D(q):=\int^\infty_0 e^{-qt} d(t,y) \df t,\qquad q>0.\] Observe that
\by
f(y):=\Exp^y\left[ \int^{\tau_0}_0 e^{-(q+\hat{r}) s } Y_s \df s \right]=q \int^\infty_0 e^{-qt} \Exp^y\left[ \int^{\tau_0 \wedge t}_0 e^{-rs} Y_s \df s\right] \df t=qD(q). 
\ey Let $r^\ast:=\hat{r}+q$ for the moment. It is not difficult to show that $f$ satisfies the ODE
\[2y^2 f''(y)+[2(\nu+1)y-1] f'(y)-r^\ast f(y)+y=0,\qquad y>0,\] with the boundary condition $f(0)=0.$ Let $\lambda(q):=\sqrt{2r^\ast+\nu^2}$ and for short $\lambda:=\lambda(0)=\sqrt{2\hat{r}+\nu^2}$.
The ODE has the following general solution
\by
f(y)&=&C_1 y^{-(\lambda(q)+\nu)/2} \exp\left(-\frac1{2y} \right)M\left(1+\frac{\lambda(q)-\nu}{2},1+\lambda(q),\frac1{2y}\right)\\
&&+C_2 y^{-(\lambda(q)+\nu)/2} \exp\left(-\frac1{2y} \right)U\left(1+\frac{\lambda(q)-\nu}{2},1+\lambda(q),\frac1{2y}\right) +\frac{r^\ast y-1}{r^\ast (r^\ast-2\nu-2)}.
\ey where $C_1$ and $C_2$ are to be determined. Note that $r^\ast> \hat{r}=2(\nu+1)+4m/\sigma^2>2(\nu+1)$. We observe that 
\be
f(y) &\le& \Exp^y\left[ \int^\infty_0 e^{-r^\ast t} Y_t I(Y_t>0) \df t \right]=\int^\infty_0 e^{-r^\ast t} \Exp^y[Y_t I(Y_t>0)] \df t\nb\\
&\le& \int^\infty_0 e^{-r^\ast t} y e^{2(\nu+1)t} \df t=\frac{y}{r^\ast-2(\nu+1)}.\qquad \label{bound}
\ee Note that as $z\rightarrow 0$,
\[U(a,b,z)\sim \frac{\Gamma(b-1)}{\Gamma(a)}z^{1-b},\qquad \Re b\ge 2.\]  Then as $y \rightarrow \infty$,
\[e^{-\frac1{2y}}U\left(1+\frac{\lambda(q)-\nu}2,1+\lambda(q),\frac1{2y}\right)y^{-(\lambda(q)+\nu)/2} \sim \frac{\Gamma(\lambda(q))}{\Gamma\left(1+(\lambda(q)-\nu)/2\right)} 2^{\lambda(q)} y^{(\lambda(q)-\nu)/2} .\] It is easy to show the fact that $r^\ast>2(\nu+1)$ implies that $(\lambda(q)-\nu)/2>1$, which means this term would not be bounded by \eqref{bound}. Hence $C_2=0.$ Second, we can use the boundary condition $f(0)=0$ to determine the coefficient $C_1$. Note that as $z\rightarrow \infty$, $|\mathrm{ph}z| \le \pi/2-\delta, a \ne 0, -1 ,\cdots,$
\[M(a,b,z) \sim \frac{\Gamma(b)}{\Gamma(a)} e^{z} z^{a-b}.\] Thus, as $y \rightarrow 0$,
\by
y^{-(\lambda(q)+\nu)/2} \exp\left(-\frac1{2y} \right)M\left(1+\frac{\lambda(q)-\nu}{2},1+\lambda(q),\frac1{2y}\right) \rightarrow \frac{\Gamma(1+\lambda(q))}{\Gamma(1+(\lambda(q)-\nu)/2)} 2^{(\lambda(q)+\nu)/2}.
\ey Since $f(0)=0$, we must have
\[C_1=\frac{\Gamma(1+(\lambda(q)-\nu)/2)}{r^\ast (r^\ast-2\nu-2)\Gamma(1+\lambda(q))} 2^{-(\lambda(q)+\nu)/2}.\] In summary,
\by f(y)&=&\frac{\Gamma(1+(\lambda(q)-\nu)/2)}{r^\ast (r^\ast-2\nu-2)\Gamma(1+\lambda(q))} (2y)^{-(\lambda(q)+\nu)/2} \exp\left(-\frac1{2y} \right)M\left(1+\frac{\lambda(q)-\nu}{2},1+\lambda(q),\frac1{2y}\right) \\
&+&\frac{ y}{r^\ast-2(\nu+1)}-\frac1{r^\ast (r^\ast-2(\nu+1))},
\ey and hence,
\by D(q)&=&\frac{\Gamma(1+(\lambda(q)-\nu)/2)}{q(\hat{r}+q) (\hat{r}+q-2\nu-2)\Gamma(1+\lambda(q))} (2y)^{-(\lambda(q)+\nu)/2} \exp\left(-\frac1{2y} \right)M\left(1+\frac{\lambda(q)-\nu}{2},1+\lambda(q),\frac1{2y}\right)\\
&+&\frac{ y}{q(\hat{r}+q-2(\nu+1))}-\frac1{q(\hat{r}+q)(\hat{r}+q-2(\nu+1))}.
\ey The three terms shall be denoted by $D_1, D_2, D_3$ respectively.
 Note that if we take the principal value (value with positive real part), then the square root function is analytic on $\C \setminus (-\infty,0]. $ The Kummer function $M(a,b,z)$ is entire in $a$ and $z$ and meromorphic in $b$ with poles at $b=-n, n=0,1,2,\cdots$. \cite[p.322]{NIST} Therefore, the Kummer function in $D$ is analytic on $\C \setminus (-\infty, -\hat{r} -\nu^2/2]$. The gamma function is meromorphic with no zeros, and with simple poles of residue $(-1)^n/n!$ at $z=-n,n=0,1,2,\cdots$. \cite[p.136]{NIST} All in all, the function $D$ is meromorphic on $\C \setminus (-\infty, -\hat{r}-\nu^2/2]$ with poles at $q=0, -\hat{r} , 2(\nu+1)-\hat{r}$ and poles where $(-\nu+\lambda(q))/2+1=-n, n=0,1,2,\cdots, $ i.e. let $k=n+1$,
\[q=2k^2-2\nu k, \qquad k < \frac{\nu}{2}, k \in \Z_+.\] 

First consider the case where $\nu \ne -1$. We deform the path of integration towards the negative $q$-axis \cite[Sections 25 and 26]{Doe} to obtain the Bromwich integral
\[\frac1{2\pi i} \int_C e^{qt} D(q) \df q =\frac1{\pi} \int^{-\infty}_{-\nu^2/2} e^{qt} \Im D(q+i0) \df q.\] Here $C$ is the path coming from $-\infty$ following the lower boundary of the cut $(-\infty,-\hat{r}-\nu^2/2]$ until $-\hat{r}-\nu^2/2$ and then returning to $-\infty$ along the upper boundary of the cut. Since $D(q+i0)$ is conjugate to $D(q-i0)$, then it simplifes to the second expression.

Note that if we let $q=-(2\hat{r}+p^2+\nu^2)/2$, then
\[\int^{-\infty}_{-\hat{r}-\nu^2/2} e^{qt} \Im D(q+i0) \df q=-\int^\infty_0 \exp\left(-\frac{(2\hat{r}+\nu^2+p^2)t}{2} \right)\Im D(-\frac{2\hat{r}+\nu^2+p^2}{2}) p \df p.\] Note that $D_2$ and $D_3$ are real-valued. Using the identity
\[M\left(\frac12+\mu-\kappa,1+2\mu,z\right)=\exp\left(\frac12 z\right) z^{-1/2-\mu} M_{\kappa,\mu}(z),\] where $M_{\kappa,\mu}(z)$ is the Whittaker function of the first kind, we rewrite $D_1$ as
\[-\frac{8}{(\nu^2+p^2)(2\hat{r}+\nu^2+p^2)(\nu^2+p^2+4\nu+4)}(2y)^{(1-\nu)/2} \exp\left(-\frac1{4y}\right) \frac{\Gamma(\frac{-\nu+\lambda}{2}+1)}{\Gamma(1+\lambda)}M_{\frac{1-\nu}{2},\frac{\lambda}{2}}\left(\frac1{2y}\right).\]
It can be shown that
\[W_{K,\mu}(z)=\frac{\pi}{\sin( 2\pi \mu)}\left[-\frac{M_{K,\mu}(z)}{\Gamma(1+2\mu)\Gamma(1/2-\mu-K)}+\frac{M_{K,-\mu}(z)}{\Gamma(1-2\mu)\Gamma(1/2+\mu-K)} \right].\] Let $K=-\kappa$ and $\mu=ip/2$. Using the fact that any real analytic function satisfies $\overline{f(\overline{z})}=f(z)$ and that $\sin ix=i\sinh x$, we obtain
\be W_{-\kappa,\frac{ip}{2}}\left(\frac1{2y}\right)=-\frac{2\pi}{\sinh(\pi p)} \Im \left( \frac{M_{-\kappa, \frac{ip}{2}}(1/(2y))}{\Gamma (1+ip)\Gamma(1-\nu/2-ip/2)} \right). \label{WIm}\ee Thus,
\[\Im\left[ \frac{M_{-\kappa,\frac{ip}{2}}\left(\frac1{2y}\right)\Gamma\left(1-\frac{\nu}{2}+\frac{ip}{2}\right)}{\Gamma(1+ip)}\right]=-\frac1{2\pi}\sinh(\pi p) W_{-\kappa, \frac{ip}{2}}\left( \frac1{2y}\right) \left|\Gamma\left(1-\frac{\nu}{2}+\frac{ip}{2}\right) \right|^2 .\]
Then we use $\Gamma(\alpha+1)=\alpha \Gamma(\alpha)$ to obtain
\by &&\frac1{2\pi i} \int_C e^{qt} D(q) \df q\\
&=&-\frac1{4\pi^2}(2y)^{(1-\nu)/2} e^{-\frac1{4y}} \int^\infty_0 \frac1{2\hat{r}+\nu^2+p^2} e^{-\frac{(2\hat{r}+\nu^2+p^2)t}{2} }\sinh(\pi p) W_{-\kappa, \frac{ip}{2}}\left( \frac1{2y}\right) \left|\Gamma\left(\frac{ip-\nu}{2}-1\right) \right|^2 p \df p.\ey

Note, however, since $D$ is meromorphic, finite many poles are placed on $\C\setminus (-\infty,-\hat{r}-\nu^2/2]$. Then we use the method of residues to identify the remaining contribution to the inverse Laplace transform. 
\by
\mathrm{Res}_{q=0} \left\{ D_2(q)-D_3(q)\right\}&=&\frac{y}{\hat{r}-2\nu-2}-\frac1{\hat{r}(\hat{r}-2\nu-2)},\\
\mathrm{Res}_{q=-\hat{r}+2\nu+2} \left\{ D_2(q)- D_3(q) \right\}&=&-\frac{y}{\hat{r}-2\nu-2} e^{-(\hat{r}-2\nu-2)t} +\frac{1}{(\hat{r}-2\nu-2)(2\nu+2)}e^{-(\hat{r}-2\nu-2)t},\\
\mathrm{Res}_{q=-\hat{r}}\left\{-D_3(q)\right\}&=&-\frac1{\hat{r}(2\nu+2)}e^{-\hat{r}t}.
\ey
We compute the residues arising from the gamma function. 
\by &&\mathrm{Res}_{q=-\hat{r}+2k^2-2\nu k}\{ e^{qt} D_1(q)\}=e^{-[\hat{r}+2k(\nu-k)]t}\frac1{2k(k-\nu)[2k(k-\nu)-\hat{r}][2k(k-\nu)-2\nu-2]}\\
&&\times \frac{(-1)^{k-1}2(\nu-2k)}{(k-1)!} (2y)^{k-\nu}\exp\left(-\frac1{2y}\right)\frac1{\Gamma(\nu-2k+1)}M\left(1-k,\nu-2k+1,\frac1{2y}\right).\ey
One can simplify the formula slightly by using the identity
\[M\left(\nu-k,\nu-2k+1,-\frac1{2y}\right)=\exp\left(-\frac1{2y}\right) M\left(-k+1,\nu-2k+1,\frac1{2y}\right).\]
The rest of resides from $D_1$ are given by
\by
\mathrm{Res}_{q=0} \{e^{qt} D_1(q)\}&=&\frac{\Gamma(1+(\lambda-\nu)/2)}{\hat{r}(\hat{r}-2\nu-2)\Gamma(1+\lambda)}(2y)^{-(\lambda+\nu)/2} e^{-\frac1{2y}}M\left(1+\frac{\lambda-\nu}2,1+\lambda,\frac1{2y}\right),\\
\mathrm{Res}_{q=-\hat{r}} \{e^{qt} D_1(q)\}&=&e^{-\hat{r}t}\frac{\Gamma(1+(|\nu|-\nu)/2)}{\hat{r}(2\nu+2)\Gamma(1+|\nu|)}(2y)^{-(|\nu|+\nu)/2} e^{-\frac1{2y}}M\left(1+\frac{|\nu|-\nu}2,1+|\nu|,\frac1{2y}\right),\\
\mathrm{Res}_{ q=-\hat{r}+2(\nu+1)}\{ e^{qt} D_1(q)\}&=&e^{-(\hat{r}-2\nu-2)t}\frac{\Gamma(1+(|\nu+2|-\nu)/2)}{(2\nu+2)(2\nu+2-\hat{r})\Gamma(1+|\nu+2|)}\\
&&\times(2y)^{-(|\nu+2|+\nu)/2} e^{-\frac1{2y}}M\left(1+\frac{|\nu+2|-\nu}2,1+|\nu+2|,\frac1{2y}\right).
\ey
Thus we collect all the residues and the inversion integral to obtain the expression for $d(t,y)$ \eqref{nune-1}.

Last, we consider the case where $\nu=-1$. Note that $q=-\hat{r}$ becomes a pole of degree 2. We rewrite $D(q)$ as
\[D(q)=\frac{\sqrt{2\pi}}{4 (r^\ast)^2 \sqrt{y}} e^{-\frac1{4y}}\left[ (2y\sqrt{1+2r^\ast} +2y+1) I_{\frac{\sqrt{1+2r^\ast}}{2}} \left( \frac1{4y} \right)+I_{\frac{\sqrt{1+2r^\ast}}2+1}\left( \frac1{4y}\right)\right]-\frac{1-r^\ast y}{(r^\ast)^2},\] where $I_\nu$ is the modified Bessel function of the first kind with the order of $\nu$. In order to calculate the residue of $e^{qt} D(q)$ at $q=-\hat{r}$, we evaluate
\by &&\left.\frac{\partial }{\partial r^\ast}\{ (r^\ast)^2 D(q)\}\right|_{r^\ast=0}\\
&=&y+\frac{\sqrt{2\pi} e^{-\frac1{4y}}}{4\sqrt{y}}\left[ 2y I_{\frac12}\left( \frac1{4y}\right)+(1+4y)\left.\frac{\partial}{\partial r^\ast} I_{\frac{\sqrt{1+2r^\ast}}2}\left( \frac1{4y}\right)\right|_{r^\ast=0}+\left.\frac{\partial}{\partial r^\ast} I_{\frac{\sqrt{1+2r^\ast}}2}\left( \frac1{4y}\right) \right|_{r^\ast=0}\right].\ey
In a manner similar to the case of $\nu=-1$ in the proof of Proposition \ref{prop:h}, we can show that
\by \left.\frac{\partial }{\partial r^\ast}\{ (r^\ast)^2 D(q)\}\right|_{r^\ast=0}=y-2y\exp\left(-\frac1{4y}\right)\sinh\left(\frac1{4y} \right)-\frac12 E_1\left( \frac1{2y}\right)=ye^{-\frac1{2y}}-\frac12 E_1\left( \frac1{2y}\right).\ey
Also note that $(\hat{r}+q)^2 D(q)|_{q=-\hat{r}}=0.$ Thus,
\[\mathrm{Res}_{q=-\hat{r}} \{e^{qt}D(q)\}=\left.\frac{\partial }{\partial q}\{ e^{qt} (\hat{r}+q)^2 D(q)\}\right|_{q=-\hat{r}}=-\frac{e^{-\hat{r}t}}{\hat{r}}\left[ye^{-\frac1{2y}}-\frac12 E_1\left( \frac1{2y}\right)\right].\] The rest of contributions to the Bromwich integral are the same as calculated in the $\nu \ne -1$ case with $\nu=-1$, which are given by
\by \mathrm{Res}_{q=0} \{e^{qt}D(q)\}&=&\frac{\hat{r}y-1}{\hat{r}^2}+\frac{\Gamma(1+\frac{\lambda+1}2)}{\hat{r}^2 \Gamma(1+\lambda)}(2y)^{-\frac{\lambda-1}2}e^{-\frac1{2y}} M\left( 1+\frac{\lambda+1}2,1+\lambda,\frac1{2y} \right)\\
\frac1{2\pi i}\int_C e^{qt} D(q) \df q&=&-\frac1{2\pi^2}y  e^{-\frac1{4y} } \int^\infty_0 \frac1{2\hat{r}+1+p^2}e^{-\frac{(1+p^2+2\hat{r})t}2 } \sinh(\pi p) W_{-1, ip/2}\left( \frac1{2y} \right) \left| \Gamma \left(\frac{ip}2-\frac12\right) \right|^2 p \df p
\ey
Collecting all above terms, we obtain the expression of $d(t,y)$ \eqref{nu=-1}.
\end{proof}

\subsection{Equivalency}
We can summarize the cash flows for the buyer (policyholder) and the seller (insurer) of the GMWB. From a buyer's standpoint, the present value of the profit (the payoff less the cost) is given by
\[e^{-rT} F_T I(\tau_0(F)>T)+\frac{w}{r}(1-e^{-rT})-G.\] From an insurer's standpoint, the present value of the profit (the payoff less the cost) is given by
\[w\int^T_{\tau_0(F)} e^{-rs} \df s I(\tau_0(F)<T)- m_w \int^{\tau_0(F)\wedge T}_0 e^{-rs} F_s \df s.\]
Note that on the insurer's income side, there is an extra parameter $m_w$ which does not appear in the policyholder's cash flows. To make the two viewpoints comparable, we first consider no expenses other than the pure cost of the GMWB, i.e. $m=m_w$, which implies all charges are used to fund the guarantee.
Observe that, unlike most financial derivatives seen in the literature, the buyer's profit from the GMWB does not exactly offset the seller's profit, which appears to be an asymmetric structure. Nevertheless, no-arbitrage theory dictates that the fair charge from a policyholder's standpoint should agree with the fair charge from an insurer's standpoint in this Black-Schole model, which otherwise would lead to an arbitrage bidding on their discrepancy.

Here we give a probabilistic proof of the equivalance of the pricing equations \eqref{pol} and \eqref{ins}. Note that \eqref{pol} can be rewritten as
\be \Exp[e^{-rT} F_T I(\tau_0(F)>T)]+w \int^T_0 e^{-rs } \df s=F_0, \label{polalt} \ee 
and \eqref{ins} can be rewritten as
\be \Exp\left[ m\int^{\tau_0(F) \wedge T}_0  e^{-rs} F_s \df s \right]= \Exp \left[  w \int^T_{\tau_0(F) \wedge T} e^{-rs } \df s \right].\label{insalt}\ee
We subtract \eqref{insalt} from \eqref{polalt} to obtain
\be \Exp \left[e^{-rT} F_T I(\tau_0(F)>T)+ \int^{\tau_0(F)\wedge T}_0 e^{-rs} (w +m F_s) \df s \right]=F_0. \label{id}\ee
It is easy to see that \eqref{pol} and \eqref{ins} are equivalent if and only if \eqref{id} holds true for all $m \in \R, T\in \R^+$. Consider the bivariate process $X_t=(t, F_t)$ and its infinitesimal generator is then given by
\[\mathcal{A} f=\frac{\partial f}{\partial t} +[(r-m)x-w] \frac{\partial f}{\partial x} +\frac{ \sigma^2}{2} \frac{\partial f^2}{\partial x^2},\qquad f=f(t,x) \in C^2_0 (\R^2).\]
Recall Dynkin's formula \cite[p.124]{Oks} that  for any stopping time $\tau$ such that $\Exp[\tau]<\infty$,
\[\Exp[f(X_\tau)]=f(x)+\Exp\left[  \int^\tau_0 \mathcal{A} f(X_s) \df s \right].\]
We let $f(t,x)=e^{-rt} x$ and $\tau=\tau_0(F)\wedge T$ and obtain
\[\Exp[e^{-r(\tau_0\wedge T)} F_{\tau_0 \wedge T}]=F_0 - \Exp\left[\int^{\tau_0 \wedge T}_0 e^{-rs} (w+mF_s )\df s  \right],\] which yields \eqref{id} after rearrangement. Therefore,  \eqref{pol} and \eqref{ins} are equivalent and the implied fair charges must be the same.

\subsection{Numerical illustration}

We provide the first numerical example to show the fair charge level determined by both points of view, which are proven to be equivalent when $m=m_w$. For the purpose of comparison, we use the same valuation basis as in \cite[p.31, Table 4]{MilSal}. The risk-free interest rate is set at $r=0.05$ and volatility coefficient is taken at two levels, $\sigma=0.2, 0.3$. A bisection method is used to determine the fair charge levels as solutions to the pricing equations. The search algorithm is programmed to reach accuracy up to 5 decimal places and then the fair charge levels, presented in Table \ref{tbl:comp}, are rounded up to the nearest basis point ($0.01\%$). In all immediate steps, we keep 10 significant digits. Each value takes about less than a minute to compute.  We used both the pricing equations \eqref{pol} and \eqref{ins} to numerically confirm the accuracy of the explicit formulas. 

\vspace{0.5cm}
\begin{table}[h] \centering
\begin{tabular}{c|c|c}
  \hline
  % after \\: \hline or \cline{col1-col2} \cline{col3-col4} ...
% \multirow{2}{*}{$w/G $}  & \multicolumn{2}{c|}{$\sigma=0.2$} & \multicolumn{2}{c}{$\sigma=0.3$}\\  %\cline{2-5}
$w/G$ & $\sigma=0.2$ & $\sigma=0.3$ \\ \hline
0.05 & 29 & 77 \\ \hline
0.06&  41  & 104 \\ \hline
0.07& 54 & 132  \\ \hline
0.08&  68 & 162  \\\hline
0.09& 82 & 192 \\\hline 
\end{tabular}
\caption{ fair charges in basis points ($m=m_w$)}
\label{tbl:comp}
\end{table}

Although within a close proximity, we found that the fair charges determined in \cite{MilSal} from the policyholder's perspective are generally overestimated, which appears to show some limitations of the numerical PDE methods introduced in the paper. The calculations based on Monte Carlo simultions would be very difficult, due to numerous recursions required to determine solutions to the pricing equations and the subsequent accumulation of estimation errors.

One should be reminded, however, that in practice only a certain percentage of the total fees is used to fund the GMWB rider, i.e. $m>m_w$. The rest of the fees and charges are used to cover overhead expenses, commisions and costs of other benefits. We explore this practical situation in a second example where $m$ differs from $m_w$. In each of the following cases, the GMWB charge is set at $80\%$ of total charges, i.e. $m_w=0.8 m.$ Then we calculate the fair charges $m_w$ from the insurer's point of view based on \eqref{ins}. The values of $m$ and $m_w$ are rounded to the nearest basis point in Table \ref{tbl:comp2}. A comparison of Table \ref{tbl:comp} and Table \ref{tbl:comp2} also shows that the total charges $m$ with $m_w=80\% m_w$ are more than $125\%$  than those with $m=m_w$. A likely explanation is that higher fees lower the value of investment account, thereby increasing the chance of early exhaustion and consequently increasing the costs of guaranteed payments from the insurer.

\vspace{0.5cm}
\begin{table}[h] \centering
\begin{tabular}{c|c|c|c|c}
  \hline
  % after \\: \hline or \cline{col1-col2} \cline{col3-col4} ...
 \multirow{2}{*}{$w/G $}  & \multicolumn{2}{c|}{$\sigma=0.2$} & \multicolumn{2}{c}{$\sigma=0.3$}\\  \cline{2-5}
& $m$ & $m_w$ & $m$ & $m_w$ \\ \hline
0.05& 37 & 29 & 101  & 81 \\ \hline
0.06& 53 & 42  & 139  &  111 \\ \hline
0.07&  71 & 56 & 179 &  143 \\ \hline
0.08&  90& 72 & 222 &  178\\\hline
0.09& 110 & 88 & 267 & 213 \\\hline 
\end{tabular}
\caption{fair charges in basis points ($m_w=0.8m$)}
\label{tbl:comp2}
\end{table}

\appendix
\section{Appendix}
\bl \label{lem:ext} When $\nu<-1$, the Laplace transform of $f$ defined in \eqref{denf} is given by
\by \int^\infty_0 e^{ -s u } f(u) \df u=(2y)^{(-\nu-\lambda)/2} \exp \left(-\frac{1}{2y}  \right)\frac{\Gamma((-\nu+\lambda)/2+1)}{\Gamma(\lambda+1)} M\left(\frac{1}{2}(-\nu+\lambda)+1, \lambda+1; \frac{1}{2y}\right),\ey for all $ \Re s\ge -\nu^2/2$ and $\lambda=\sqrt{2s+\nu^2}$.
\el

\begin{proof}
It follows from \eqref{WIm} that
\by
W_{-\kappa, ip/2}\left(\frac1{2y}\right)\left| \Gamma\left(-\frac{\nu}{2}+\frac{ip}{2}\right)  \right|^2 \sinh(\pi p)=2\pi \Im\left( \frac{M_{-\kappa, ip/2}(1/(2y))\Gamma(-\nu/2+ip/2)}{\Gamma(1+ip)(ip/2+\nu/2)} \right).
\ey It is known in \cite[p.95,(10)]{BucGerman} that
\be M_{\kappa, iq}(x)\sim x^{1/2+iq},\qquad 0 <q \rightarrow +\infty.\qquad \label{magM}\ee We also known from \cite[(5.11.9)]{NIST} that for fixed $x \in \R$,
\be |\Gamma(x+iy)|\sim \sqrt{2\pi} \exp\left(-\frac{\pi}{2} y\right)y^{x-1/2},\qquad 0<y\rightarrow +\infty. \qquad \label{magG}\ee Using \eqref{magM} and \eqref{magG}, we obtain that as $p \rightarrow +\infty$,
\[ \left|W_{-(1-\nu)/2,ip/2}\left(\tfrac{1}{2y}\right)\Gamma\left(\tfrac{-\nu+ip}{2}\right)^2 \sinh(\pi p) p\right|\le C_1 \exp(\pi p/4) p^{\nu/2-1}, \] where $C_1$ is a positive constant.
Arguing as in the proof of Watson's lemma~(c.f. \cite{Olv}), it follows that 
\begin{equation}\label{eq4}
 |f(u)|\le C_2 u^{-3/2} e^{-\nu^2 u/2} \quad \text{as $u\to+\infty$} .
\end{equation}
The positive constant $C_2$ may depend on $\nu$ and $y$ but is independent of (large) $u$.
We conclude from \eqref{eq4} that the Laplace transform
\[ F(s)=\int_0^\infty e^{-su} f(u)\,du \]
is analytic for $\Re s>-\nu^2/2$ and continuous for $\Re s\ge -\nu^2/2$ (since $u^{-3/2}$ is integrable at $\infty$.) Let $G$ be the expression for the Laplace transform of $\tau_0$ determined for $s>0$ in \eqref{Laptau}. Recall that the principle value of the square root function $\lambda(s)=\sqrt{2s+\nu^2}$ is analytic for $\Re s>-\nu^2/2$ and the Kummer function $M(a,b,z)$ is analytic in $a$ and $b$ except for $b=0,-1,-2,\cdots$. Thus $G$ is analytic for $\Re s>-\nu^2/2$. Since $F$ and $G$ agree on the positive real axis, they must agree for all $\Re s> -\nu^2/2.$ The expression of $G$ holds for $\Re s=-\nu^2/2$ due to the continuity of $F$.
\end{proof}

\bibliography{refs-feng}
\end{document}